\documentclass[a4paper,11pt]{article}
\pdfoutput=1

\usepackage{jheppub} 
\usepackage{lineno}
\usepackage{physics}

\usepackage[dvipsnames]{xcolor}

\newcommand{\Omegab}{E_b}
\newcommand{\delm}{\Delta m}

\renewcommand{\textit}[1]{#1}

\usepackage[normalem]{ulem} 

\usepackage[most]{tcolorbox} 
\usepackage{lipsum} 

\newtcolorbox{plan}{
  enhanced jigsaw,     
  breakable,           
  pad at break=1mm,    
  colback=gray!15,     
  colframe=gray!30,    
  boxrule=0pt,         
  arc=2mm,             
  left=2mm,right=2mm,  
  top=1mm,bottom=1mm,
  before skip=6pt, after skip=6pt,
}

\title{One Sum To Rule Them All: A Second Order Master Rate Sum Rule for Charm Decays}

\author[a]{Margarita Gavrilova}
\author[b]{Yuval Grossman}
\author[b]{Guglielmo Papiri}
\author[c]{Stefan Schacht}

\affiliation[a]{Walter Burke Institute for Theoretical Physics, California Institute of Technology, 1200 E. California
Boulevard, Pasadena CA 91125, USA}
\affiliation[b]{Department of Physics, LEPP, Cornell University, Ithaca, NY 14853, USA}
\affiliation[c]{Institute for Particle Physics Phenomenology, Department of Physics,
Durham University, Durham DH1 3LE, United Kingdom}

\emailAdd{mgavr@caltech.edu}
\emailAdd{yg73@cornell.edu}
\emailAdd{gp343@cornell.edu}
\emailAdd{stefan.schacht@durham.ac.uk}

\subheader{\hfill \textnormal{IPPP/26/16}}

\abstract{
We show that within the Standard Model any system of hadronic weak charm decays related by $U$-spin satisfies the following rate sum rule:
\begin{equation*} 
\frac{\text{(sum of CF and DCS CKM-free rates)}} {\text{(sum of SCS CKM-free rates)}} = 1\,, 
\end{equation*}
which holds up to second order in $U$-spin breaking.
We test this sum rule against available data and find that it is well satisfied in all cases. For systems in which some decay rates have not yet been measured, we use this sum rule to predict the missing rates. 
}

\begin{document}
\maketitle
\flushbottom

\section{Introduction}\label{sec:intro}

Calculating charm decay rates from first principles is notoriously difficult. However, approximate flavor symmetries of QCD can be used to derive relations among decay rates (and other observables) of charm decays. Crucially, such predictions are independent of a model of hadronization, so confronting them with data can shed light on the underlying strong dynamics and provide tests of physics beyond the Standard Model (SM).

Flavor-symmetry relations are based on an expansion in a symmetry-breaking parameter, and thus their predictive power hinges on how well this expansion is behaved for the observables under consideration. There are examples where the symmetry limit works very well. Yet, there are some cases where the data indicate large symmetry breaking.

In some cases the origin of the breaking effect is understood, while in others it remains unclear. For example, large deviations at the level of rates can arise from kinematic effects even when the underlying amplitude-level breaking is moderate. A classic example is the ratio~\cite{PDG:2024cfk}
\begin{align}
\frac{\mathcal{B}(\phi \rightarrow K^+K^-)}{\mathcal{B}(\phi\rightarrow K_L K_S)} &\sim 1.5\,,
\end{align}
which exhibits a deviation from the isospin limit of one by about 50\%. 
The majority of this effect can be accounted for by the relevant phase space factor that scales like the momenta of the outgoing kaon to the third power~\cite{Bramon:2000qe}. 
Another example is the ratio of $B_s$ to $B_d$ production fraction ratios $f_s/f_d \sim 1/4$, which differs from its symmetry-limit expectation of one by about a factor of four~\cite{PDG:2024cfk, LHCb:2021qbv, ATLAS:2015esn, CMS:2022wkk}. 
A qualitative understanding of  this effect can be obtained in the Lund model~\cite{Andersson:1983ia}.

On the other hand, there are cases where the data hints at large breaking and where we currently do not have an understanding of the underlying source. One example is the $2.7\sigma$ ``$U$-spin anomaly'' in the direct CP asymmetries of $D^0\to K^+K^-$ and $D^0\to \pi^+\pi^-$~\cite{LHCb:2022lry}
\begin{align}
\frac{a_{CP}^{\mathrm{dir}}(D^0\rightarrow \pi^+\pi^-)}{a_{CP}^{\mathrm{dir}}(D^0\rightarrow K^+K^-)} \sim +3
\end{align}
vs.~the $U$-spin limit expectation of $-1$ (for theoretical interpretations see Refs.~\cite{Schacht:2022kuj,Bause:2022jes}).
Another recent example can be found in two-body charmless $B$ decays, where SU(3)$_F$ breaking at the level of $\mathcal{O}(10)$ has been reported in Refs.~\cite{Bhattacharya:2022akr,Berthiaume:2023kmp,Bhattacharya:2025wcq}. Other works, however, find smaller SU(3)$_F$ breaking to be compatible with the data~\cite{Huber:2021cgk,BurgosMarcos:2025xja}.

These examples motivate an effort to develop a better understanding of flavor-symmetry breaking. In practice, this leads to two complementary goals. First, we would like to construct relations that go beyond the symmetry limit and are therefore expected to receive parametrically smaller corrections if the expansion is well behaved, while offering a more stringent test when it is not. Second, confronting such higher-order relations with data provides an additional diagnostic test of the expansion itself, helping to assess when and how flavor symmetry can be used reliably in phenomenological analyses.

In the case of $U$-spin, symmetry breaking is controlled by the quark-mass difference
\begin{equation}
\delm\equiv m_s - m_d\,.
\end{equation}
This parameter is dimensionful, so defining a dimensionless parameter that controls the expansion requires an additional energy scale. We therefore introduce a presumably small dimensionless parameter schematically as
\begin{equation}\label{eq:epsilon_sch}
\varepsilon \equiv \frac{\delm}{\text{(some energy scale)}}\,.
\end{equation}
It is common to take the relevant scale to be of order $\Lambda_{\rm QCD}$, which suggests a typical breaking of order $20\%$. This is a crude assumption, motivated primarily by the ratio of decay constants of the kaon and pion, $f_K/f_\pi-1\sim 0.2$. However, the effective scale controlling a given observable can depend on the process, QCD dynamics, and kinematics, and can differ significantly from $\Lambda_{\rm QCD}$. In this work we therefore adopt an agnostic approach. Although we quote our results in powers of $\varepsilon$, we always have in mind that the only well-defined parameter that controls the breaking is $\Delta m$, and that $\varepsilon$ is not guaranteed to be small. Thus,  for example, a correction of $\mathcal{O}(\varepsilon)$ should be understood as a correction proportional to $\Delta m$. In the present work, we do not attempt to estimate the effective scale in the denominator.

Going beyond leading order provides an additional handle on the expansion: if the symmetry is a good approximation and the breaking is indeed controlled by a small parameter, relations that hold at second order should exhibit smaller deviations from the symmetry limit than relations with linear corrections. A well-known example of a set of decays related by $U$-spin (a ``$U$-spin system''), where the expansion has been tested, is given by $D^0\to P^+P^-$ decays, with $P=K,\pi$. This system has been studied extensively; see, for example, Refs.~\cite{Brod:2012ud, Grossman:2019xcj, Muller:2015lua, Grossman:2006jg, Hiller:2012xm, Grossman:2013lya, Grossman:2012ry, Fleischer:2025zhl}. The two relations that hold in the $U$-spin symmetry limit and are theoretically broken by first-order corrections can be expressed as ratios of CKM-normalized rates, $\hat{\Gamma}$, defined precisely in Eqs.~\eqref{eq:Gamma_hat_def}--\eqref{eq:f_CKM}. The symmetry-limit prediction for these ratios is unity, whereas experimentally one finds
\begin{equation}\label{eq:DtoPP_LO_intro}
    \frac{\hat{\Gamma}(D^0 \to K^+ K^-)}{\hat{\Gamma}(D^0 \to \pi^+ \pi^-)}
    = 2.81 \pm 0.06 \, , \qquad
    \frac{\hat{\Gamma}(D^0 \to K^+ \pi^-)}{\hat{\Gamma}(D^0 \to \pi^+ K^-)}
    = 1.21\pm 0.02\,,
\end{equation}
while the relation that is unity in the symmetry limit and theoretically holds through second order in the symmetry breaking~\cite{Grossman:2012ry} is, experimentally,
\begin{equation}\label{eq:DtoPP_NLO_intro}
    \frac{\hat{\Gamma}(D^0 \to K^+ \pi^-) + \hat{\Gamma}(D^0 \to \pi^+ K^-)}
         {\hat{\Gamma}(D^0 \to K^+ K^-) + \hat{\Gamma}(D^0 \to \pi^+ \pi^-)}
    = 0.84\pm 0.01\,.
\end{equation}
Empirically, although one of the ratios in Eq.~\eqref{eq:DtoPP_LO_intro} exhibits a large deviation from the symmetry-limit value $1$, the second-order relation in Eq.~\eqref{eq:DtoPP_NLO_intro} is satisfied much better.
This illustrates the practical value of second-order sum rules: if a second-order sum rule is satisfied better than the first-order sum rules, this points to a well-behaved expansion and supports treating $\varepsilon$ as a small parameter.

Higher-order sum rules have been explored before, though the existing literature remains limited: second-order rate sum rules for selected charm-decay systems were derived in Ref.~\cite{Grossman:2012ry}, and general constructions of higher-order amplitude sum rules were discussed in Refs.~\cite{Gavrilova:2022hbx,Gavrilova:2024npn}. Higher-order rate sum rules for $b$ baryon decays have been found in Ref.~\cite{Wen:2025ibn}. In this paper we present general tools for constructing second-order relations between decay rates, and apply them to charm decays. In particular, we show that the following master sum rule is universal for arbitrary systems of weak charm decays:
\begin{equation}\label{eq:master-intro}
    \frac{\text{(sum of CF and DCS CKM-free rates)}}
         {\text{(sum of SCS CKM-free rates)}}
    = 1 + \mathcal{O}\!\left(\varepsilon^2\right)\,.
\end{equation}
Here, CF, SCS, and DCS denote Cabibbo-favored, singly Cabibbo-suppressed, and doubly Cabibbo-suppressed channels, respectively.  
CKM-free rates refer to rates where the respective CKM factors are divided out, see Sec.~\ref{sec:charm}. Depending on the group-theoretical structure of a given system, additional second-order sum rules may exist. Some of them follow from the construction described in this paper, but not all of them. A systematic approach to deriving the complete set of higher-order sum rules between rates for arbitrary systems (not limited to charm decays) and at all orders beyond second order is left for an upcoming publication~\cite{Gavrilova:inpreparation}.

This paper is organized as follows. In Section~\ref{sec:theory} we present the general framework and derive the second-order master sum rule formulas, Eqs.~\eqref{eq:sym_shmu_integer}--\eqref{eq:sym_shmu_half_integer}. In Section~\ref{sec:master} we apply these results to charm decays and obtain the universal charm master sum rule, Eq.~\eqref{eq:master-intro}. In Section~\ref{sec:examples} we consider many applications to hadronic charm decays and present comparisons to data where available. We find that, in all the examples where the master sum rule in Eq.~\eqref{eq:master-intro} can be tested, it is well satisfied, despite the fact that some of the leading-order sum rules appear to be badly broken. For systems where only partial data is currently available, we use the master sum rule in Eq.~\eqref{eq:master-intro} to set limits on the unmeasured branching fractions.
In Section~\ref{sec:conclusions}, we conclude. Additional derivations and comments are collected in Appendices~\ref{app:shmu}-\ref{app:two-body}.

\section{Generalities of second order $SU(2)_F$ rate sum rules}\label{sec:theory}

This section develops the theoretical ingredients that we use to obtain second-order $U$-spin rate sum rules for weak charm decays. At the same time, most of the statements we make are substantially more general than this final application.
Motivated by this generality, we formulate the discussion in a way that applies to any $SU(2)_F$ flavor symmetry, including isospin, $U$-spin, and $V$-spin. Throughout this section, $a$ and $b$ denote the two light flavors related by $SU(2)_F$: for isospin they correspond to $u$ and $d$, for $U$-spin to $d$ and $s$, and for $V$-spin to $u$ and $s$.
We define
\begin{equation}
m_\text{av} = {m_a+ m_b \over 2}, \qquad
\Delta m\equiv m_b-m_a,
\qquad
\Delta m^2\equiv (m_b-m_a)^2\,,
\end{equation}
where $m_a$ and $m_b$ are the quark masses.
The symmetry limit corresponds to the symmetric point $m_a=m_b=m_\text{av}$, while symmetry breaking is controlled by the quark-mass difference $\Delta m$. 

In general, there may be additional sources of symmetry breaking. In particular, for isospin and $V$-spin, the quarks in the doublet carry different electric charges. In the following, we do not discuss this effect explicitly, and instead comment at the end why the argument extends to arbitrary sources of symmetry breaking.

This section is organized as follows. In Section~\ref{sec:symmetry_argument} we present the general symmetry argument: we show that any symmetry-limit relation that is symmetric under $a\leftrightarrow b$ exchange automatically persists to second order in $SU(2)_F$ breaking. In Section~\ref{sec:shmu} we describe a method that provides symmetry-limit linear relations among observables proportional to squared amplitudes. Section~\ref{sec:identical} discusses how to interpret this method in the presence of identical multiplets. Finally, in Section~\ref{sec:sym+shmu=love} we combine the symmetry argument with the construction in Section~\ref{sec:shmu} to derive second-order sum rules. The key output is the set of master equations in Eqs.~\eqref{eq:sym_shmu_integer} and~\eqref{eq:sym_shmu_half_integer}. Comments on the applicability and limitations of the various assumptions are given in the corresponding subsections.

\subsection{Symmetry argument for second order $SU(2)$ sum rules}\label{sec:symmetry_argument}

In this section we formulate an argument that allows us to identify $SU(2)_F$ symmetry-limit relations among physical observables that automatically persist to second order in
the symmetry breaking. We treat the observables as functions of the quark masses $(m_a,m_b)$ and assume that they are analytic in
the neighborhood of the symmetric point. Although in nature the quark masses have fixed values, here we treat $(m_a,m_b)$ as continuous parameters.

At the fundamental level, any deviation from the exact $SU(2)_F$ limit originates from the quark-mass difference $\Delta m$. We therefore study symmetry breaking by expanding the observables around the symmetric point $m_a=m_b=m_\text{av}$  in powers of $\Delta m$. Of course, $\Delta m$ is dimensionful, so the notion of it being ``small'' is not well defined and requires an additional energy scale. Here we do not attempt to
identify this scale or to estimate the corresponding dimensionless expansion parameter, since it can depend on the specific process,
QCD dynamics, and kinematics (see the discussion in the introduction). Nevertheless, $\Delta m$ is the only well-defined parameter that
quantifies the departure from the exact symmetry limit, and we thus use it as an expansion parameter.
Thus in what follows, ``second order'' refers to corrections that start at $\mathcal{O}(\Delta m^2)$ in the expansion about $m_a=m_b=m_\text{av}$, where $\Delta m^2 \equiv (m_b - m_a)^2$. The idea is that, if symmetry breaking is indeed small, then at least formally one can introduce an energy scale and define a dimensionless parameter $\varepsilon\propto \Delta m$. The deviation from the symmetry limit at the level of observables is then controlled by $\varepsilon$, so that observables admit a power-series expansion in $\varepsilon$, and $\mathcal{O}(\Delta m^2)$ is equivalent to $\mathcal{O}(\varepsilon^2)$. For the formal discussion in this section we keep the expansion in powers of $\Delta m$, while for most of this paper we assume that the expansion is well behaved and present our results in powers of $\varepsilon$.

Finally, we note in passing that although the scope of this work is sum rules between observables, the symmetry argument in this section also applies at the level of amplitudes. At the amplitude level, this has been proven in full generality (with generalizations to higher orders in symmetry breaking) in Ref.~\cite{Gavrilova:2022hbx}, and corresponds to the statement that ``$s$-type amplitude sum rules valid in the symmetry limit automatically hold to second order in symmetry breaking.'' See the discussion in Sec.~III.C of Ref.~\cite{Gavrilova:2022hbx}.

\paragraph{General symmetry argument.}
The symmetry argument in this section is essentially the statement that the Taylor expansion of a function
symmetric under $m_a\leftrightarrow m_b$ exchange contains no term linear in $(m_b-m_a)$ when expanded about the
symmetric point $m_a=m_b$.

We now elaborate on the argument. Consider an abstract symmetry-limit relation among physical observables related by $SU(2)_F$, which we denote by 
\begin{equation}\label{eq:S_sym_limit}
    \mathcal{S}(m_a,m_b)=0
    \qquad\text{at}\qquad
    m_a=m_b=m_\text{av}\,,
\end{equation}
or equivalently,
\begin{equation}\label{eq:S_sym_limit_2}
    \mathcal{S}(m_\text{av},m_\text{av})=0\,.
\end{equation}
At this stage, $\mathcal{S}(m_a,\,m_b)$ can be any linear or nonlinear expression built from physical observables
(for example, decay rates, scattering cross sections, CP asymmetries, etc.). We also do not specify the nature of the observables, which may arise from weak or strong processes. When $\mathcal{S}(m_a,\,m_b)$ is linear
in the observables, it is standard to refer to Eq.~\eqref{eq:S_sym_limit} as a (symmetry-limit) sum rule. We make one assumption about $\mathcal{S}(m_a,\,m_b)$ and impose one additional property:
\begin{enumerate}
\item We assume that $\mathcal{S}(m_a,\,m_b)$ can be treated as a function of $(m_a,m_b)$ and that it is analytic in the
neighborhood of the symmetric point $m_a = m_b =m_\text{av}$.
\item We restrict our attention to relations $\mathcal{S}$ that are symmetric under the exchange of the quark masses,
\begin{equation}\label{eq:S_exchange_sym}
        \mathcal{S}(m_a,m_b)=\mathcal{S}(m_b,m_a)\,.
    \end{equation}
\end{enumerate}
Given this, we expand $\mathcal{S}$ around the symmetric point by writing
\begin{equation}\label{eq:ma_mb_delta}
    m_a=m_\text{av}-\frac{\Delta m}{2}\,,\qquad
    m_b=m_\text{av}+\frac{\Delta m}{2}\,,
\end{equation}
and, using analyticity,
\begin{align}\label{eq:S_expand_delta}
    \mathcal{S}\left(m_\text{av}-\frac{\Delta m}{2},m_\text{av}+\frac{\Delta m}{2}\right)
    &=
    \mathcal{S}\left(m_\text{av},m_\text{av}\right)
    +\frac{\Delta m}{2} \left.\left(\frac{\partial \mathcal{S}}{\partial m_b}-\frac{\partial \mathcal{S}}{\partial m_a}\right)\right|_{(m_\text{av},m_\text{av})}
    +\mathcal{O}(\Delta m^2)\,.
\end{align} 
Since Eq.~\eqref{eq:S_sym_limit} implies $\mathcal{S}(m_\text{av},m_\text{av})=0$, the leading correction is naively controlled by
the linear term. However, the exchange symmetry that we impose in Eq.~\eqref{eq:S_exchange_sym} implies that
\begin{equation}
    \mathcal{S}\left(m_\text{av}-\frac{\Delta m}{2},m_\text{av}+\frac{\Delta m}{2}\right)
    =
    \mathcal{S}\left(m_\text{av}+\frac{\Delta m}{2},m_\text{av}-\frac{\Delta m}{2}\right)\,,
\end{equation}
so $\mathcal{S}(m_\text{av}-\frac{\Delta m}{2},m_\text{av}+\frac{\Delta m}{2})$ is an even function of ${\Delta m}$. Therefore the coefficient of the
linear term in Eq.~\eqref{eq:S_expand_delta} must vanish, and we conclude that
\begin{equation}\label{eq:S_Odelta2}
    \mathcal{S}(m_a,m_b)=\mathcal{O}\left(\Delta m^2\right)\,.
\end{equation}
In other words, any symmetry-limit relation (linear or non-linear) 
$\mathcal{S}(m_a,m_b)=0$ that is symmetric under $m_a\leftrightarrow m_b$
automatically persists to second order in $SU(2)_F$ breaking.

We note that the above argument can be generalized to any source of symmetry breaking. As long as the observable is invariant under exchange of $a$ and $b$ and is analytic, the leading non-vanishing corrections are second order in the breaking parameters.

\paragraph{Recipe for second-order sum rules.}
We now specialize the general argument above to sum rules, {\it i.e.},~relations linear in observables.
Consider a collection of $SU(2)_F$-related observables $\{{\mathcal{O}}_\alpha\}$ for channels labeled by $\alpha$. (We only consider observables that are finite in the symmetry limit.)
As above, we treat each $\mathcal{O}_\alpha$ as a function of $(m_a,m_b)$, analytic near $m_a=m_b=m_\text{av}$. We restrict our attention to systems (and thus sets of observables) for which one can define, for each channel $\alpha$,
an $SU(2)_F$-conjugate channel $\overline{\alpha}$ obtained by exchanging the two flavors $a\leftrightarrow b$ in all external states and in the Hamiltonian, such that the corresponding observables satisfy
\begin{equation}\label{eq:sigma_conj_condition}
    \mathcal{O}_{\overline{\alpha}}(m_a,m_b)=\mathcal{O}_\alpha(m_b,m_a)\,.
\end{equation}
Whenever Eq.~\eqref{eq:sigma_conj_condition} holds, we can form symmetric combinations
\begin{align}\label{eq:Sigma_def}
    \Sigma_\alpha(m_a,m_b)&\equiv \mathcal{O}_\alpha(m_a,m_b)+\mathcal{O}_{\overline{\alpha}}(m_a,m_b)\nonumber\\
    &= \mathcal{O}_\alpha(m_a,m_b)+\mathcal{O}_\alpha(m_b,m_a)\,,
\end{align}
which are manifestly symmetric under $m_a\leftrightarrow m_b$. Any symmetry-limit sum rule that involves only symmetric combinations,
\begin{equation}\label{eq:sym_sumrule_limit}
    \sum_\alpha c_\alpha\,\Sigma_\alpha(m_\text{av},m_\text{av})=0\,,
\end{equation}
defines an $\mathcal{S}$ of the form
\begin{equation}\label{eq:S_from_Sigma}
    \mathcal{S}(m_a,m_b)\equiv \sum_\alpha c_\alpha\,\Sigma_\alpha(m_a,m_b)\,,
\end{equation}
which satisfies $\mathcal{S}(m_\text{av},m_\text{av})=0$ and $\mathcal{S}(m_a,m_b)=\mathcal{S}(m_b,m_a)$.
Therefore, by the general argument above,
\begin{equation}\label{eq:sym_sumrule_2nd}
    \sum_\alpha c_\alpha\,\Sigma_\alpha(m_a,m_b)=\mathcal{O}(\Delta m^2)\,.
\end{equation}
In other words, the practical recipe is: identify observables for which an $SU(2)_F$ conjugate exists and satisfies
Eq.~\eqref{eq:sigma_conj_condition}, form $\Sigma_\alpha$, and then any symmetry-limit sum rule that can be written purely in terms of $\Sigma_\alpha$
automatically persists up to $\mathcal{O}(\Delta m^2)$ corrections.

Before commenting on the applicability of this recipe, we briefly discuss the question of the small parameter $\varepsilon$.
Given an observable $\mathcal{O}(m_a,m_b)$, we expand it around the symmetric point $m_a=m_b=m_\text{av}$,
\begin{equation}\label{eq:sigma_expand_Deltam}
    \mathcal{O}(m_a,m_b)
    = \mathcal{O}(m_\text{av},m_\text{av})
    + \frac{\Delta m}{2}\left.\left(\frac{\partial \mathcal{O}}{\partial m_b}-\frac{\partial \mathcal{O}}{\partial m_a}\right)\right|_{(m_\text{av},m_\text{av})}
    + \mathcal{O}(\Delta m^2)\,.
\end{equation}
If the expansion is well behaved, then from here it is natural to estimate the  energy scale that controls the expansion as
\begin{equation}\label{eq:Lambda_def}
    \Lambda \sim
    \frac{\mathcal{O}(m_\text{av},m_\text{av})}
    {\left.\left(\frac{\partial \mathcal{O}}{\partial m_b}-\frac{\partial \mathcal{O}}{\partial m_a}\right)\right|_{(m_\text{av},m_\text{av})}}\,.
\end{equation}
Although in general we do not have a way to calculate \(\Lambda\), the condition \(\Delta m \ll \Lambda\) provides a formal criterion for when the expansion is well behaved.

\paragraph{Comments on applicability.}
The recipe for deriving second-order sum rules between observables described above relies on two assumptions. 

\begin{enumerate}
    \item {\it The system is complete under the flavor exchange $a\leftrightarrow b$.}
    We assume that whenever a channel belongs to the system, the channel obtained by exchanging the two $SU(2)_F$ flavors
    $a\leftrightarrow b$ also belongs to the system. Even when all channels are kinematically allowed, this property is not automatic:
    
    \begin{itemize}
    \item {\it Strong processes (any $SU(2)_F$).}
    For purely strong scattering and decays, this property always holds.

    \item {\it Weak processes related by $U$-spin.}
    For $U$-spin systems, the exchange $d\leftrightarrow s$ preserves electric charge, so all the $U$-spin-related channels are allowed.

    \item {\it Weak processes related by isospin (or $V$-spin).}
    For purely hadronic external states, the exchange $u\leftrightarrow d$ (or $u\leftrightarrow s$) can map a channel to one
    that is forbidden by $U(1)_{\rm em}$, so some of the $SU(2)_F$-related channels are absent. This issue is avoided in systems that include additional $SU(2)_F$ singlets, such as leptons, which can balance the total electric charge. We refer the reader to Ref.~\cite{Gavrilova:2022hbx} for details, in particular the discussion around Eq.~(2.19) and the example in Sec.~V.C.2 therein.
    \end{itemize}

    \item {\it The chosen observables satisfy the conjugation condition.}
    The observables must be chosen such that Eq.~\eqref{eq:sigma_conj_condition}
    holds.
\begin{itemize}
\item For strong observables (for example, differential or integrated cross sections and rates), this is always the case.
\item For weak observables (for example, differential or integrated decay rates and cross sections, CP asymmetries), CKM factors generally
spoil Eq.~\eqref{eq:sigma_conj_condition}.
However, when a single $SU(2)_F$ irrep dominates the effective Hamiltonian, one can instead work with CKM-free observables,
obtained schematically by dividing the physical observable by its CKM factor, for example $\hat{\Gamma}=\Gamma/\text{CKM}$, where $\Gamma$ is a decay rate and $\hat{\Gamma}$ is the corresponding CKM-free rate. We discuss this construction in detail in Sec.~\ref{sec:charm}.
\item We also comment on a  standalone strong observable, namely a hadron mass. Hadron masses can be interpreted as matrix elements of the Hamiltonian in the hadron rest frame between one-hadron states and thus are given by $h\to h$ amplitudes, where $h$ is a hadron. As noted above, at the amplitude level the symmetry argument has been understood in Ref.~\cite{Gavrilova:2022hbx} including generalizations to arbitrary orders in the symmetry breaking.
\end{itemize}
\end{enumerate}

\subsection{The Shmushkevich method}\label{sec:shmu}

In simple terms, the Shmushkevich method~\cite{Shmushkevich:1955,DushinShmushkevich:1956} derives relations between processes related by isospin. It states that once all but one of the isospin indices are summed over, the result is the same for any value of the remaining unsummed index.

The Shmushkevich method did not gain much attention in the flavor-physics community, presumably since it was originally developed and applied in the context of nuclear scattering. In this section we review the method and reformulate it in a general $SU(2)_F$ language, emphasizing that its applicability extends well beyond its original setting and can be used as a powerful tool for deriving symmetry-limit sum rules between observables that depend quadratically on amplitudes.

The Shmushkevich method was developed as a clever procedure for deriving symmetry-limit isospin relations among hadronic scattering cross sections without performing an explicit Clebsch--Gordan decomposition of amplitudes. In modern language, one works in the isospin limit, where the relevant interaction Hamiltonian transforms as a singlet under $SU(2)_I$, and the external states are given by direct products of arbitrary $SU(2)_I$ irreps, and uses the symmetry to obtain linear relations among physical channels. A systematic presentation of the method is given, for example, in Ref.~\cite{PinskiMacfarlaneSudarshan:1965}, and examples of its applications to exclusive and inclusive hadronic cross sections can be found in Refs.~\cite{PinskiMacfarlaneSudarshan:1965,LipkinPeshkin1972Inclusive,Kyriakopoulos1974IsospinRelations}. Ref.~\cite{LipkinPeshkin1972Inclusive} also points out applications to $U$-spin. More general formulations and extensions to $SU(3)$ and other symmetry groups are discussed in Ref.~\cite{MacfarlaneMukundaSudarshan:1964}.

The Shmushkevich method was originally derived for hadronic scattering via strong interaction in the isospin limit.  In this work we extend it to a wider class of physical systems including weak decays related by $U$-spin. First, if one follows the derivation in Ref.~\cite{PinskiMacfarlaneSudarshan:1965}, it is straightforward to see that the method applies to any set of strong scattering processes related by any $SU(2)_F$, since the derivation relies only on $SU(2)$ symmetry and does not use any assumptions specific to isospin. Second, the universality argument of Ref.~\cite{Gavrilova:2022hbx} ensures that all the results also hold for decays mediated by the weak Hamiltonian. Indeed, moving irreps between the external states and the Hamiltonian does not change the group-theoretic structure of the underlying amplitude sum rules, and therefore does not change the resulting relations among rates or cross sections. In particular, introducing a non-singlet irrep in the Hamiltonian, or replacing a two-body scattering process by a decay, does not change the group-theoretic problem: in all cases, the only input that matters is the set of irreps present in the system.

The Shmushkevich method is a powerful symmetry-limit tool, but, as implemented in the literature, it does not provide a clear path toward incorporating symmetry-breaking effects. However, combined with the symmetry argument in Section~\ref{sec:symmetry_argument}, it allows us to identify symmetry-limit sum rules that persist to second order. In what follows we describe the method and later apply the symmetry argument to the resulting sum rules to obtain the second-order relations.

\paragraph{Shmushkevich's master equation.} To emphasize the power of Shmushkevich's result, we present it here in an abstract language for an arbitrary $SU(2)_F$, and in terms of generic observables that depend quadratically on amplitudes. In doing so, we further narrow the class of observables that we consider. While in Section~\ref{sec:symmetry_argument} the observables $\mathcal{O}$ were arbitrary functions of amplitudes that are analytic in the quark masses and satisfy the conjugation condition in Eq.~\eqref{eq:sigma_conj_condition}, we now restrict ourselves to a subset of such observables, denoted $\hat{\sigma} \propto \abs{A}^2$, that depend quadratically on the amplitudes.

The key technical inputs of the derivation are the orthogonality and completeness of Clebsch--Gordan coefficients: by forming suitable sums over the $SU(2)_F$ $m$-quantum numbers, one eliminates the group-theory factors and obtains a set of linear master equations among cross sections. We refer the reader to Ref.~\cite{PinskiMacfarlaneSudarshan:1965} for the derivation in the special case of strong scattering cross sections related by isospin, and to Appendix~\ref{app:shmu} for the derivation restated in the general language of $SU(2)_F$ representation theory. Here we simply state the assumptions and the result, and provide a brief discussion.

Consider a system of processes related by an $SU(2)_F$ symmetry such that the group-theoretical structure of the system is given by a direct product of $r$ arbitrary distinguishable $SU(2)_F$ irreps 
\begin{equation}\label{eq:gen_system}
    I_1 \otimes I_2 \otimes \dots \otimes I_r\,.
\end{equation}
Here ``distinguishable'' means that the irreps correspond to different particle multiplets, or to identical multiplets that are distinguished by their momenta. We discuss the question of identical versus distinguishable irreps further in Section~\ref{sec:identical}.
Additionally, we assume that all processes in the system allowed by $SU(2)_F$ can be realized.

Each process in the system can be labeled by listing $r$ $m$-quantum numbers (QNs), one for each irrep in Eq.~\eqref{eq:gen_system}. We use $\hat{\sigma}(m_1,\,m_2,\,\dots,\, m_r)$ to denote a generic observable, defined as a CKM-free amplitude squared convolved with the appropriate kinematic function. Examples of such observables are strong or CKM-free weak scattering cross sections and decay rates. We emphasize that $\hat{\sigma}$ is not a function of the $m$-QNs, rather, the $m_j$ are labels that distinguish different processes/channels of the system. We also emphasize once more that the notion of observable used here is narrower than in Section~\ref{sec:symmetry_argument}, as we require that the observables are proportional to squared amplitudes, $\hat{\sigma}\propto \abs{A}^2$. This proportionality is important since, strictly speaking, the Shmushkevich method yields relations at the level of squared amplitudes. However, in the symmetry limit, where hadrons within a given $SU(2)_F$ multiplet are degenerate and share identical kinematics, these relations carry over straightforwardly to the corresponding physical cross sections and rates.

Following Ref.~\cite{PinskiMacfarlaneSudarshan:1965}, we define
\begin{equation}\label{eq:sigma_i_def}
    \hat{\sigma}_i(m_i) \equiv \sum_{\{m_j\}_{j\neq i}} \hat{\sigma}(m_1,\dots,m_r)\,,
\end{equation}
where the sum runs over $m_j=-I_j,\dots,I_j$ for all $j\neq i$, while $m_i$ is held fixed and such that the corresponding processes are allowed by $SU(2)_F$. We refer to $\hat{\sigma}_i(m_i)$ as an inclusive observable in the sense that it sums over all channels in the system with a fixed value of the $m$-QN of irrep $I_i$.

The Shmushkevich master equation (see, {\it e.g.,} Eq.~(11) of Ref.~\cite{PinskiMacfarlaneSudarshan:1965} and Eq.~(\ref{app:Master}) in Appendix~\ref{app:shmu} of the present paper) states that, in the exact $SU(2)_F$ symmetry limit, $\hat{\sigma}_i(m_i)$ is independent of $m_i$, i.e.\ the inclusive cross section obtained by fixing the $m$-QN of a single irrep and summing over all others is the same for all $m_i$. Explicitly, Shmushkevich's master equation yields $2I_i$ sum rules
\begin{equation}\label{eq:shmu_master}
    \hat{\sigma}_i(-I_i) = \hat{\sigma}_i(-I_i+1) = \dots = \hat{\sigma}_i(I_i)\,.
\end{equation}
Equalities such as
\begin{equation}\label{eq:shmu_trivial}
    \hat{\sigma}_i(m_i)=\hat{\sigma}_i(-m_i)
\end{equation}
are trivial in the sense that they follow directly from the symmetry under the simultaneous sign flip of all $m$-QNs (equivalently, a flavor exchange $a\leftrightarrow b$),
\begin{equation}\label{eq:full_flip}
    \hat{\sigma}(m_1,\dots,m_r)=\hat{\sigma}(-m_1,\dots,-m_r)\,.
\end{equation}
For higher irreps with $I_i>1/2$, Eq.~\eqref{eq:shmu_master} gives additional, non-trivial relations among the inclusive observables $\hat{\sigma}_i(m_i)$ beyond the simple $m_i\leftrightarrow -m_i$ exchange. Concretely, it yields $I_i$ additional independent equalities for integer $I_i$, and $I_i-\tfrac12$ independent equalities for half-integer $I_i$.

We note that the Shmushkevich master sum rules in Eq.~\eqref{eq:shmu_master} do not guarantee a complete set of linearly independent symmetry-limit sum rules for an arbitrary system. In fact, there are examples where additional sum rules exist. 
Such relations can be understood within a more general framework, which we will discuss in an upcoming publication~\cite{Gavrilova:inpreparation}.

\subsection{The Shmushkevich method in the presence of identical multiplets}\label{sec:identical}

The Shmushkevich sum rules in Eq.~\eqref{eq:shmu_master} are derived under the assumption that all the multiplets in the system, Eq.~\eqref{eq:gen_system}, are distinguishable, see the explicit derivation in Appendix~\ref{app:shmu}. This means that these sum rules can be applied directly in the following cases:
\begin{enumerate}
    \item {\it Differential observables} (i.e.\ differential cross sections and differential rates). Even when identical irreps are present, in the differential case each external multiplet carries an implicit momentum label, so the multiplets are effectively distinguishable. For example, if the final state contains two doublets $P^+=(K^+,\,\pi^+)$, then at a fixed point in phase space it is understood as containing $P^+_{p_1}P^+_{p_2}$. Thus a final state with $\pi^+(p_1)K^+(p_2)$ is distinct from the final state with $\pi^+(p_2)K^+(p_1)$, and both momentum assignments appear as separate terms in sum rules of the form in Eq.~\eqref{eq:shmu_master}.
    \item {\it Integrated observables} (i.e.\ total cross sections and decay rates) in systems where all irreps in the final state are distinct, i.e. correspond to different particle multiplets. We typically define integrated observables with fixed momenta of all initial-state particles (in the case of a decay the decaying particle is taken to be at rest in its center-of-mass frame) so as discussed above then all irreps in the initial state are effectively distinguishable from each other and from the irreps in the final state. The irrep in the Hamiltonian does not carry a momentum label but is clearly distinguishable from all the particle multiplets. This is why it is sufficient to require that the irreps in the final state are distinguishable: if no particle multiplet appears more than once among the final-state multiplets, the momentum-labeled sum rule in Eq.~\eqref{eq:shmu_master} can be integrated directly, yielding a relation among integrated observables with no additional combinatorial factors.

\end{enumerate}

In the presence of identical irreps in the final state, \emph{i.e.},~irreps that correspond to the same particle multiplet, translating momentum-labeled sum rules into relations among integrated physical observables can, in general, introduce combinatorial factors. In what follows, we explain how to account for these factors. We then compute them explicitly in the special case where the Shmushkevich equation is written for an irrep that is distinguishable from all other irreps in the system. We show that, for this class of Shmushkevich sum rules, the combinatorial factors cancel even when identical irreps are present in the final state, and the simple form of Eq.~\eqref{eq:shmu_master} is preserved for integrated observables.

We consider an $SU(2)$ system with the generic group-theoretical structure in Eq.~\eqref{eq:gen_system}. Suppose that there are $g$ distinct irreps $I_1,\dots,I_g$ in the system, and that the irrep $I_i$ appears $k_i$ times, $i=1,\dots,g$, that is, the system has the structure
\begin{equation}\label{eq:gen_system_ident}
    \underbrace{I_1\otimes\cdots\otimes I_1}_{k_1}\otimes
    \underbrace{I_2\otimes\cdots\otimes I_2}_{k_2}\otimes\cdots\otimes
    \underbrace{I_g\otimes\cdots\otimes I_g}_{k_g}\,.
\end{equation}
Note that when the irrep $I_i$ is distinguishable from all other irreps in the system, $k_i=1$. As pointed out above, this is the case for all irreps in the initial state, the irrep in the Hamiltonian, and any irrep in the final state that corresponds to a distinct particle multiplet. To begin with, we interpret Eq.~\eqref{eq:shmu_master} as a relation between differential observables. To make this explicit, we rewrite Eq.~\eqref{eq:shmu_master} as
\begin{equation}\label{eq:shmu_dif}
    \hat{\sigma}^{\text{diff}}_i(-I_i) = \hat{\sigma}^{\text{diff}}_i(-I_i+1) = \dots = \hat{\sigma}^{\text{diff}}_i(I_i)\,,
\end{equation}
where $\hat{\sigma}^{\text{diff}}_i(m_i)$ has an implicit dependence on the momenta of all external particles. Integrating Eq.~\eqref{eq:shmu_dif} over the final state phase space then gives a corresponding relation among integrated quantities,
\begin{equation}\label{eq:shmu_int}
    \int d\Pi\, \hat{\sigma}^{\text{diff}}_i(-I_i) = \int d\Pi\, \hat{\sigma}^{\text{diff}}_i(-I_i+1) = \dots = \int d\Pi\, \hat{\sigma}^{\text{diff}}_i(I_i)\,,
\end{equation}
where $\int d \Pi$ is an integral over all final state momenta without any symmetry factors. In the symmetry limit this step is straightforward (recall that Eq.~\eqref{eq:shmu_master} is written in the symmetry limit), since all channels share the same kinematics and therefore the same integration domain. In the case where all irreps in the final state are distinguishable, each individual term in the sums in Eq.~\eqref{eq:shmu_int} corresponds to an integrated observable,
\begin{equation}\label{eq:sigma_phys_dist}
    \text{no identical irreps:}\quad\int d\Pi\, \hat{\sigma}^{\text{diff}}(m_1,\dots,m_r)
    = \hat{\sigma}^{\rm int}(m_1,\dots,m_r)\,,
\end{equation}
and thus the integrals over the inclusive sums of the differential $\hat{\sigma}^{\text{diff}}_i(m_i)$ map directly into inclusive sums of integrated observables,
\begin{equation}
    \text{no identical irreps:}\quad \int d\Pi\, \hat{\sigma}^{\text{diff}}_i(m_i)
    = \hat{\sigma}^{\rm int}_i(m_i)\,.
\end{equation}
The question that we ask is how the integrals $\int d\Pi\, \hat{\sigma}^{\text{diff}}_i(m_i)$ map into integrated observables when identical multiplets are present in the final state.

There are two sources of combinatorial factors. First, in the presence of identical particles the integrated observables are defined with the usual symmetry factors in the phase space integral that modify Eq.~\eqref{eq:sigma_phys_dist}. To describe these factors it is enough to focus on one group of $k$ identical irreps $I$ in the final state.

Any physical state in this group can be specified by listing how many particles in this physical state correspond to each component of the multiplet $I$. That is, we specify the multiplicities $\{n_m\}$, with $m=-I,-I+1,\dots,I$, where $n_m$ counts how many of the $k$ particles are in the component with a given $m$-QN. By construction these multiplicities satisfy
\begin{equation}
\sum_{m=-I}^{I} n_m = k\,.
\end{equation}
For example, consider a group of $k = 3$ identical copies of the $I=\tfrac12$ doublet of pseudoscalar mesons $P^+=(K^+,\pi^+)$ in the final state. If the particular final state of interest is $K^+K^+\pi^+$, then two particles are in the $K^+$ component and one particle is in the $\pi^+$ component. In terms of multiplicities, this corresponds to $n_{+1/2}=2$ and $n_{-1/2}=1$, and we have $n_{+1/2} + n_{-1/2} = k$.

When integrating over phase space, the momentum-labeled expression $\hat{\sigma}^{\text{diff}}$ overcounts configurations that differ only by permutations of identical particles. For the fixed physical channel specified by $\{n_m\}$, the corresponding symmetry factor is 
\begin{equation}\label{eq:sym_factor}
S_{\{n_m\}} \equiv \prod_{m=-I}^{I} n_m!\,,
\end{equation}
so that Eq.~\eqref{eq:sigma_phys_dist} is modified to
\begin{equation}\label{eq:sigma_phys_int}
\int d\Pi\,\hat{\sigma}^{\text{diff}}(m_1,\dots,m_r)
= S_{\{n_m\}}\;\hat{\sigma}^{\rm int}(m_1,\dots,m_r)\,,
\end{equation}
where we keep all other external states implicit. In the above example of the $K^+K^+\pi^+$ final state, the symmetry factor is $S_{\{n_m\}} = 2! \cdot 1 = 2$. If the system contains several groups of identical irreps in the final state, the same construction applies to each group: one specifies multiplicities $\{n_m^I\}$ for each irrep $I$, and the total symmetry factor is the product of the corresponding factors for all groups.

The second source of combinatorial factors is that many different momentum assignments to the particles within the final state, several of which may appear in a single $\hat{\sigma}^{\text{diff}}_i(m_i)$, can correspond to the same physical channel after integration. For example, consider a decay whose final state contains the three mesons $K^+K^+\pi^+$. The following ordered momentum assignments may all appear in one $\hat{\sigma}^{\text{diff}}_i(m_i)$:
\begin{equation}\label{eq:KKpi-mom}
    K^+(p_1)K^+(p_2)\pi^+(p_3)\,,\quad
    K^+(p_1)\pi^+(p_2)K^+(p_3)\,, \quad \text{and} \quad
    \pi^+(p_1)K^+(p_2)K^+(p_3).
\end{equation}
Therefore, for each $\hat{\sigma}^{\text{diff}}_i(m_i)$ one must count how many distinct momentum-labeled terms map, upon integration, onto the same physical integrated observable.

While identifying these terms is straightforward in practice, the most general counting is somewhat cumbersome. We will not go into the details of this fully general counting. Instead, we focus on cases where the counting is simple and leads to particularly simple form of sum rules among integrated observables. In particular, in the case that we consider below the complete sums over external $m$-QNs in $\hat{\sigma}_i(m_i)$ automatically include all possible momentum assignments, and the resulting multiplicity factors cancel with the usual symmetry factors discussed above.

The simplified counting applies to sum rules in which the irrep $I_i$ used to write Eq.~\eqref{eq:shmu_dif} has multiplicity $k_i=1$. To illustrate the counting in this case, we again focus on one group of $k$ identical irreps $I$. In the differential description the $k$ copies of the irrep $I$ are distinguished by their momenta and therefore we can think of them as ordered slots. For a fixed physical state specified by $\{n_m\}$, there are
\begin{equation}\label{eq:Nnm}
N_{\{n_m\}} = \frac{k!}{\prod_{m=-I}^{I} n_m!} = \frac{k!}{S_{\{n_m\}}}
\end{equation}
distinct ordered assignments of the $m$-components of irrep $I$ to these $k$ slots and thus $N_{\{n_m\}}$ distinct differential cross sections that result in the same physical observable after integration. For the example above with $K^+K^+\pi^+$, we have
\begin{equation}
N_{\{n_m\}}=\frac{3!}{2!\,1!}=3\,,
\end{equation}
corresponding to the three ordered momentum assignments in Eq.~\eqref{eq:KKpi-mom}.

The key to the cancellation advertised above, given that the irrep $I_i$ used to write the Shmushkevich equation has $k_i = 1$, is that if one of the $N_{\{n_m\}}$ terms appears in $\hat{\sigma}_i(m_i)$, then all of them appear, since the sums in $\hat{\sigma}^{\text{diff}}_i(m_i)$ run independently over the full range $m=-I,\dots,I$ for each momentum-labeled slot (irrep). All of these differential terms therefore map, after integration, to the same physical observable in Eq.~\eqref{eq:sigma_phys_int}. For a physical state specified by $\{n_m\}$, this means that $\hat{\sigma}^{\rm int}$ appears $N_{\{n_m\}}$ times in the integrated sum, and each appearance comes with the symmetry factor $S_{\{n_m\}}$. Using Eq.~\eqref{eq:Nnm}, the resulting combinatorial factor is
\begin{equation}
    N_{\{n_m\}}\,S_{\{n_m\}} = k!\,.
\end{equation}
Thus, for each group of $k$ identical irreps one obtains an overall factor of $k!$. For a system with $g$ groups as in Eq.~\eqref{eq:gen_system_ident}, the total factor is $k_1!\cdots k_g!$, and since this factor is the same for all channels it can be canceled from the master equation. Concretely, Eq.~\eqref{eq:shmu_int} becomes
\begin{equation}\label{eq:shmu_phys_fact}
    (k_1!\cdots k_g!)\,\hat{\sigma}^{\rm int}_i(-I_i)
    = (k_1!\cdots k_g!)\,\hat{\sigma}^{\rm int}_i(-I_i+1)
    = \dots
    = (k_1!\cdots k_g!)\,\hat{\sigma}^{\rm int}_i(I_i)\,,
\end{equation}
so the factorial factors cancel and we arrive at
\begin{equation}\label{eq:shmu_phys}
    \hat{\sigma}^{\rm int}_i(-I_i) = \hat{\sigma}^{\rm int}_i(-I_i+1) = \dots = \hat{\sigma}^{\rm int}_i(I_i)\,.
\end{equation}
For a concrete example, see Appendix~\ref{app:identical_DPPP}, where we work out explicitly a three-body decay system with two identical doublets in the final state.

We thus conclude that Shmushkevich's master equation retains its form for integrated physical observables even when identical irreps are present in the final state, provided it is written for an irrep $I_i$ with multiplicity $k_i = 1$. The absence of additional combinatorial factors is a consequence of the inclusive sums in Eq.~\eqref{eq:shmu_master} together with the above condition on $I_i$, and it does not hold in general. In particular, it is easy to see from the example in Eq.~\eqref{eq:KKpi-mom} that if one writes the Shmushkevich master equation for the doublet $P^+_{p_1}$, for example, the equation equates two inclusive sums such that one side includes in the sum the first two momentum assignments, while the other includes only the third (with the remaining particles and their momenta kept fixed). In this case, the cancellation does not take place.

\subsection{Second order master sum rules}\label{sec:sym+shmu=love}

In this section we bring together the symmetry argument of Section~\ref{sec:symmetry_argument} and the Shmushkevich method for symmetry-limit sum rules discussed in Sections~\ref{sec:shmu} and~\ref{sec:identical} to construct second-order sum rules between observables proportional to amplitudes-squared. The key idea is to take symmetry-limit sum rules and form linear combinations that are manifestly symmetric under the $a\leftrightarrow b$ exchange. By the symmetry argument, all such sum rules then hold up to corrections of order $\mathcal{O}(\varepsilon^2)$, where the meaning of the corrections is as discussed in Section~\ref{sec:symmetry_argument}.

We take as our starting point the Shmushkevich master equation in Eq.~\eqref{eq:shmu_master}, which for any irrep $I$ in the system encodes $2I$ symmetry-limit sum rules. Of course, any linear combination of sum rules is again a sum rule, so in principle one could start from any basis of symmetry-limit sum rules to follow the program outlined above. However, for sum rules written in the form of Eq.~\eqref{eq:shmu_master}, the linear combinations that are symmetric under flavor conjugation are immediately evident.

In particular, for any integer irrep $I> \tfrac12$ we can write $I$ sum rules that follow from Eq.~\eqref{eq:shmu_master} as
\begin{equation}\label{eq:sym_shmu_integer}
    \hat{\sigma}_i(I_i) + \hat{\sigma}_i(-I_i)
    =
    \hat{\sigma}_i(I_i-1) + \hat{\sigma}_i(-I_i+1)
    =
    \dots
    =
    2\hat{\sigma}_i(0)\,.
\end{equation}
Similarly, for any half-integer irrep $I>\tfrac12$ we can write $I-\tfrac12$ sum rules that follow from Eq.~\eqref{eq:shmu_master} as
\begin{equation}\label{eq:sym_shmu_half_integer}
    \hat{\sigma}_i(I_i) + \hat{\sigma}_i(-I_i)
    =
    \hat{\sigma}_i(I_i-1) + \hat{\sigma}_i(-I_i+1)
    =
    \dots
    =
    \hat{\sigma}_i(+\tfrac12) + \hat{\sigma}_i(-\tfrac12)\,.
\end{equation}
Equations~\eqref{eq:sym_shmu_integer} and~\eqref{eq:sym_shmu_half_integer} are linear combinations of the symmetry-limit sum rules in Eq.~\eqref{eq:shmu_master} and are manifestly symmetric under $SU(2)_F$ conjugation.
Thus they hold to second order. Therefore, for any higher irrep in the system one automatically obtains $I$ (for integer $I$) or $I-\tfrac12$ (for half-integer $I$) second-order sum rules of the form in Eqs.~\eqref{eq:sym_shmu_integer}--\eqref{eq:sym_shmu_half_integer}.

In the next sections we apply this result to the special case of weak charm decays. Before we proceed, we note that since the Shmushkevich method is not guaranteed to yield all symmetry-limit sum rules, the master sum rules in Eqs.~\eqref{eq:sym_shmu_integer} and~\eqref{eq:sym_shmu_half_integer} are likewise not guaranteed to yield all second-order sum rules for a system under consideration.

\section{The Master Sum Rule for Charm Decays}\label{sec:master}

In this section we apply the general results of Section~\ref{sec:theory} to the phenomenologically relevant case of hadronic weak charm decays, that is those mediated by $c \to u q \bar q'$ with $q,q'=s,d$. We do not consider semi-leptonic decays mediated by $c \to q \ell \bar \nu$. We begin by summarizing the $U$-spin-limit effective Hamiltonian for charm decays in Section~\ref{sec:charm} and by defining the relevant observables, namely integrated CKM-free decay rates, which satisfy the conditions of Section~\ref{sec:theory}. In Section~\ref{sec:master-charm} we apply the second-order master sum rule of Section~\ref{sec:sym+shmu=love} to the triplet of the charm Hamiltonian and derive the universal second-order master sum rule for weak charm decays. The key results are Eqs.~\eqref{eq:master_ratio_second_order} and~\eqref{eq:master}.

\subsection{Charm Hamiltonian and the CKM-free rates}\label{sec:charm}

\paragraph{The two generation approximation.}
In what follows, we make one well-justified approximation: we neglect the effects of the third generation when considering charm decay rates.
In practice, this is done by assuming that the CKM matrix is a unitary $2 \times 2$ matrix, parametrized by a single parameter $\theta_C$. Using $\lambda \equiv \sin\theta_C$, where $\lambda$ is the Wolfenstein parameter, it is given by
\begin{equation}\label{eq:VCKM-approx}
V_{\text{CKM}}^{(2)} =
\begin{pmatrix}
\cos\theta_C & \sin\theta_C \\
-\sin\theta_C & \cos\theta_C
\end{pmatrix}
=
\begin{pmatrix}
\sqrt{1-\lambda^2} & \lambda \\
-\lambda & \sqrt{1-\lambda^2}
\end{pmatrix}.
\end{equation}
The unitarity of $V^{(2)}_{\rm CKM}$ implies that, in this approximation, the CKM factor in front of the $\Delta U = 0$ component of the charm Hamiltonian vanishes, and thus only the $\Delta U = 1$ component contributes.

The above approximation is valid up to relative corrections of order $\lambda^4\sim 10^{-3}$. This is well below the current experimental precision for all relevant rates as well as the expected second-order $U$-spin--breaking effects. We therefore adopt this approximation throughout and keep the associated $\mathcal{O}(\lambda^4)$ uncertainty implicit.

\paragraph{Hamiltonian for charm decays.} 

Using the notation of Ref.~\cite{Gavrilova:2022hbx}, and working in the two-generation approximation described above, 
the $U$-spin-limit effective Hamiltonian for charm decays transforms as a triplet and can be written as
\begin{equation}\label{eq:Heff-charm}
    \mathcal{H}_{\text{eff}}^{\text{$U$-spin limit}} = \sum_{m = -1}^{1} f_{1,m} H^1_m\,,
\end{equation}
where $H^1_m$ denotes a component of a triplet with a given $m$ quantum number, and $f_{1,m}$ denotes the corresponding CKM-factor dependence. We have
\begin{equation} \label{H1}
        H^1_{1} =  (\bar{u} s) (\bar d c),\qquad
        H^1_{-1}= -(\bar{u} d) (\bar s c), \qquad
        H^1_0 =  {(\bar{u} s) (\bar s c)-(\bar{u} d) ( \bar d c)\over \sqrt{2}},
\end{equation}
and
\begin{align}
f_{1,1} &= V_{cd}^* V_{us} = -\lambda^2, \\
 f_{1,-1} &= -V_{cs}^* V_{ud}= -(1-\lambda^2), \\
  f_{1,0} &= \frac{V_{cs}^* V_{us} - V_{cd}^* V_{ud}}{\sqrt{2}}
 = \sqrt{2} \left(\lambda \sqrt{1-\lambda^2}\right)\,, \label{eq:CKM}
\end{align}
where in the last equalities we use the two-generation approximation in Eq.~\eqref{eq:VCKM-approx}. Note the factor of $\sqrt{2}$ in the definition of $f_{1,0}$, which is necessary to ensure the correct normalization of the operator $H^1_0$. With these conventions, $H^1_1$ is the Hamiltonian for doubly Cabibbo-suppressed (DCS) charm decays, $H^1_{-1}$ is the Hamiltonian for Cabibbo-favored (CF) charm decays, and $H^1_0$ is the Hamiltonian for singly Cabibbo-suppressed (SCS) charm decays. 

At leading order, the $U$-spin-limit Hamiltonian for charm decays transforms as a triplet ($u_H=1$). We also take all particles in the initial and final states (\emph{i.e.},~the external states) to transform as irreducible $U$-spin representations (``pure irreps'' in the language of Refs.~\cite{Gavrilova:2022hbx,Gavrilova:2024npn}). With this setup, any system of charm decays has the following group-theoretical structure under $U$-spin:
\begin{equation}\label{eq:charm-structure}
    \left(u_1 \otimes \dots \otimes u_r\right)\otimes u_H \,,\qquad u_H = 1\,,
\end{equation}
where $u_1,\,\dots,\,u_r$ denote the $U$-spin representations of the external states in the system, and we label representations by their total $U$-spin.

\paragraph{CKM-free rates.}
The discussion in Section~\ref{sec:theory} was kept very general: in Section~\ref{sec:symmetry_argument} we considered generic observables $\hat{\sigma}$ that satisfy Eq.~\eqref{eq:sigma_conj_condition}, and in Sections~\ref{sec:shmu}--\ref{sec:sym+shmu=love} we additionally required that the observables depend quadratically on amplitudes, i.e.\ $\hat{\sigma}\propto \abs{A}^2$. From now on we focus on a specific class of observables relevant for weak charm decays that satisfies Eq.~\eqref{eq:sigma_conj_condition} and is proportional to squared amplitudes, namely integrated CKM-free decay rates. We denote them by $\hat{\Gamma}(i\to f)$, where we use $i$ to label an initial charm hadron and $f$ is a collection of final state particles. We define the CKM-free rates, in a way analogous to CKM-free amplitudes in Ref.~\cite{Gavrilova:2022hbx}, as total decay rates of charmed hadrons $\Gamma(i\rightarrow f)$ normalized by the corresponding CKM-factor $f_\text{CKM}$, that is
\begin{equation}\label{eq:Gamma_hat_def}
    \hat \Gamma (i\rightarrow f) \equiv \frac{\Gamma(i \rightarrow f)}{ f_\text{CKM}}\,,
\end{equation}
where
\begin{equation}\label{eq:f_CKM}
f_\text{CKM} = 
\begin{cases}
f_{\mathrm{CF}} = \abs{V_{cs}V_{ud}}^{2}\ = (1-\lambda^2)^2\,, & \\[4pt]
f_{\mathrm{SCS}} = \abs{V_{cs}V_{us}}^{2}\ = \abs{V_{cd}V_{ud}}^{2}\ = \lambda^{2}(1-\lambda^2)\,, & \\[4pt]
f_{\mathrm{DCS}} = \abs{V_{cd}V_{us}}^{2}\ = \lambda^{4}\,. & 
\end{cases}
\end{equation}

Note that for SCS rates, we intentionally leave out the factor of $2$ that one would obtain by squaring $f_{1,0}$ in Eq.~\eqref{eq:CKM}. This is a purely conventional choice: we adopt it so that the master sum rule for charm decays takes the simple form given in the introduction, Eq.~\eqref{eq:master-intro}, and the CKM factors are aligned with the convention commonly used in the literature.

\subsection{Second order master sum rule in charm}\label{sec:master-charm}

As discussed in Section~\ref{sec:master}, for each integer irrep $I>1/2$ there exist at least $I$ second-order sum rules, given in Eq.~\eqref{eq:sym_shmu_integer}, while for each half-integer irrep $I>1/2$ there are at least $I-1/2$ second-order sum rules, given in Eq.~\eqref{eq:sym_shmu_half_integer}. Given the approximations of Section~\ref{sec:charm}, the $U$-spin-limit charm Hamiltonian transforms as a triplet ($u_H=1$). Therefore, all of weak charm decay systems have the group-theoretical structure in Eq.~\eqref{eq:charm-structure}. Thus, according to Section~\ref{sec:sym+shmu=love}, for any such charm system there exists at least one second-order sum rule that follows from Eq.~\eqref{eq:sym_shmu_integer} when applied to the triplet in the Hamiltonian, i.e.\ with $I_i$ set to $u_H=1$ in Eq.~\eqref{eq:sym_shmu_integer}. This is the universal charm master sum rule that we derive below. We note that, while all systems of weak charm decays have this sum rule, additional sum rules may exist. In particular, a sufficient condition for the existence of additional second-order sum rules is the presence of higher representations among the external states. We present an example of such a sum rule in Section~\ref{sec:neutral_charm_baryon}.

We now apply the second-order master sum rule in Eq.~\eqref{eq:sym_shmu_integer} to the triplet in the Hamiltonian. Labeling the $U$-spin related rates by listing the $m$-QNs of the external states and the Hamiltonian as $\hat{\Gamma}(m_1,\dots,m_r,m_H)$, we define the inclusive CKM-free rate with fixed Hamiltonian $m$-QN by
\begin{equation}\label{eq:Gamma_inclusive_H}
\hat{\Gamma}_H(m_H)\equiv \sum_{\{m_j\}_{j\neq H}} \hat{\Gamma}(m_1,\dots,m_r,m_H)\,,
\end{equation}
where the sum runs over $m_j=-u_j,\dots,u_j$ for all external irreps $u_1,\,\dots,\,u_r$, Eq.~\eqref{eq:sym_shmu_integer} gives
\begin{equation}\label{eq:GammaH_master}
    \hat{\Gamma}_H(1)+\hat{\Gamma}_H(-1)
    = 2\,\times \frac{1}{2}\times \hat{\Gamma}_H(0) + \mathcal{O}(\varepsilon^2)\,.
\end{equation}
The factor of $2$ is the one that appears in Eq.~\eqref{eq:sym_shmu_integer}.
The explicit factor of $\tfrac{1}{2}$ reflects the need to match the group-theoretic object that
enters the Shmushkevich equation. Shmushkevich's master equation relates inclusive observables for different components
$m_H$ of the triplet in the Hamiltonian. For $m_H=\pm1$ the inclusive CKM-free rate
$\hat{\Gamma}_H(\pm1)$ is normalized to the corresponding component in Eq.~\eqref{eq:Heff-charm}.
For $m_H=0$, however, the Hamiltonian component carries an extra normalization. From Eq.~\eqref{eq:CKM} the correct normalization is
\begin{equation}
    |f_{1,0}|^2\simeq 2\,|V_{cs}V_{us}|^2\,.
\end{equation}
Since our definition of CKM-free SCS rates in Eqs.~\eqref{eq:Gamma_hat_def}--\eqref{eq:f_CKM} divides by $|V_{cs}V_{us}|^2$ (rather than by
$|f_{1,0}|^2$), the quantity $\hat{\Gamma}_H(0)$ differs by a factor of $2$ from the correctly normalized $m_H=0$
component that appears in the Shmushkevich equation. Equivalently, the Shmushkevich input for the $m_H=0$ component is
$\tfrac{1}{2}\,\hat{\Gamma}_H(0)$, which is why Eq.~\eqref{eq:GammaH_master} contains the explicit product $2\times\tfrac{1}{2}$.

Finally, we present the second order sum rule in Eq.~\eqref{eq:GammaH_master} in a form of a ratio as
\begin{equation}\label{eq:master_ratio_second_order}
    R_H \equiv \frac{\hat{\Gamma}_H(1)+\hat{\Gamma}_H(-1)}{\,\hat{\Gamma}_H(0)}
    = 1+\mathcal{O}(\varepsilon^2)\,.
\end{equation}
As discussed in Sec.~\ref{sec:charm}, in charm $m=1$ corresponds to DCS decays, $m=-1$ to CF decays, and $m=0$ to SCS decays, so Eq.~\eqref{eq:master_ratio_second_order} has the structure of Eq.~\eqref{eq:master-intro},
\begin{equation}\label{eq:master}
    \frac{\left(\text{sum of CF and DCS CKM-free rates}\right)}
         {\left(\text{sum of SCS CKM-free rates}\right)}
    = 1 + \mathcal{O}\left(\varepsilon^2\right) \, .
\end{equation}
This is the second order $U$-spin master sum rule for charm decays.

\section{Examples}\label{sec:examples}

In this section we discuss several concrete charm decay systems and write the corresponding second-order charm master sum rules of the form in Eq.~\eqref{eq:master} explicitly in terms of CKM-free rates. Where data is available, we confront the $U$-spin--breaking expansion with measurements. In the examples for which all relevant branching ratios are measured (Sections~\ref{sec:DtoPP}--\ref{sec:DtoPPP}), we find that the second-order master sum rules are satisfied more accurately than relations that theoretically receive $\mathcal{O}(\varepsilon)$ corrections. Where only partial data exists, we assume that $U$-spin--breaking expansion is well behaved and derive limits on unmeasured branching ratios that follow from the second order charm master sum rule.

In the following, we restrict our attention to singly charmed hadrons that are stable under strong and electromagnetic interactions and therefore decay exclusively via the weak interaction. These are the ground-state charm hadrons containing a single charm quark. In total, this set consists of three charmed mesons and four charmed baryons. Their basic properties, including quark content, masses, and lifetimes, are summarized in Table~\ref{tab:numerical}.
Out of the 7 hadrons, six are part of a multiplet that does not decay strongly. The $\Omega_c$ is the exception, as the rest of its multiplet do decay strongly.

\begin{table}[t]
\centering
\begin{tabular}{|c|c|c|c|c|}
\hline  & & & &\\[-13pt]
Hadron & Quark content & $|U,m_U\rangle$ & Mass [MeV] & Lifetime [$10^{-13}$ s] \\[1pt]
\hline
 & & & &\\[-13pt]
$D^0$        & $c \bar u$ & $\lvert 0,0\rangle$   & $1864.84\pm0.05$ & $4.103 \pm 0.010$ \\
$D^+$        & $c\bar d$ & $\lvert \tfrac12,-\tfrac12\rangle$ & $1869.66\pm0.05$ & $10.33 \pm 0.05$ \\
$D_s^+$      & $c\bar s$ & $\lvert \tfrac12,+\tfrac12\rangle$ & $1968.35\pm0.07$ & $5.012 \pm 0.022$ \\[2pt]
\hline & & & &\\[-13pt]
$\Lambda_c^+$ & $cud$ & $\lvert \tfrac12,+\tfrac12\rangle$   & $2286.46\pm0.14$ & $2.026 \pm 0.010$ \\
$\Xi_c^+$     & $cus$ & $\lvert \tfrac12,-\tfrac12\rangle$ & $2467.71\pm0.23$ & $4.53 \pm 0.05$ \\
$\Xi_c^0$     & $cds$ & $\lvert 0,0\rangle$ & $2470.44\pm0.28$ & $1.504 \pm 0.028$ \\
$\Omega_c^0$  & $css$ & $\lvert 1,-1\rangle$  & $2695.3\pm0.4$ & $2.73 \pm 0.12$ \\
\hline
\end{tabular}
\caption{Ground-state singly charmed hadrons that decay exclusively via the weak interaction. The $U$-spin quantum numbers are given in the $\lvert U,m_U\rangle$ basis, where $U$-spin acts on the $(d,s)$ quark doublet. Masses and lifetimes are taken from the PDG~\cite{PDG:2024cfk}. A comprehensive list of the isospin, $U$-spin and $V$-spin wave functions of charmed baryons can be found in Ref.~\cite{Adolph:2022ujd}.}
\label{tab:numerical}
\end{table}

For the final state, we only consider particles belonging to multiplets that are irreducible representations of $U$-spin. This excludes, for example, multiplets containing neutral mesons which are linear combinations of several such representations such as the $\eta$ and~$\eta'$.

\paragraph{Experimental input.} For our numerical results, we use branching ratios, lifetimes, and CKM parameters as reported by the PDG~\cite{PDG:2024cfk}. 
For the mixing angle defined in Eq.~\eqref{eq:VCKM-approx}, we are using
\begin{equation}
\sin \theta_C = \lambda = 0.22501\pm0.00068\,.
\end{equation}

The second-order master sum rule, Eq.~\eqref{eq:master}, is written in terms of CKM-free decay rates, defined in Eq.~\eqref{eq:Gamma_hat_def}, which can be expressed in terms of measured quantities as
\begin{equation}\label{eq:Gammahat_from_data}
    \hat{\Gamma}(i\to f) \equiv \frac{\Gamma(i \rightarrow f)}{ f_\text{CKM}} =\frac{\mathcal{B}(i\to f)}{\tau_i\, f_{\rm CKM}}\,,
\end{equation}
where $\mathcal{B}(i\to f)$ is the branching fraction of the $i\to f$ decay, $\tau_i$ is the lifetime of the decaying particle $i$, and $f_{\rm CKM}$ is the corresponding CKM factor defined in Eq.~\eqref{eq:f_CKM}.

Note that since we neglect CP violation in our analysis, we use CP-averaged branching ratios throughout. For bookkeeping, we organize the branching ratios into $U$-spin pairs and label the channels within each pair as ``decay'' and ``conjugate decay.'' This assignment is a convention: for each system, we choose a particular $U$-spin doublet in the initial or final state and define the pairs relative to that choice.  For example, if the defining doublet is $P=(P_1,P_2)$, we refer to channels with $P_2$ as ``decays'' and to those with $P_1$ as ``conjugate decays''.

In our numerical analysis we compute the uncertainties of our numerical results by determining their $\Delta\chi^2$-profile. In doing so we make several numerical approximations:
\begin{itemize}
\item
We use the experimentally extracted central values for the Cabibbo angle and for the lifetimes and neglect their small experimental errors.
\item
We do not take correlations between the branching-ratio measurements into account.
\item
Whenever asymmetric uncertainties are quoted in the experimental data, we symmetrize the errors by re-centering the quoted central value to the midpoint of the corresponding confidence interval and reporting a symmetric uncertainty equal to half of the interval width. In our numerical analysis, when the resulting uncertainty is asymmetric, we quote a symmetrized uncertainty in the final results. We checked that the difference of the results from ones obtained using Gaussian error propagation are negligible compared to the respective errors. 
\end{itemize}
These approximations have only a mild impact on the numerical results and do not affect our qualitative conclusions.

\paragraph{Presentation of the results.} 
For each system with available data, we first test the symmetry-limit sum rules. In principle, there are many ways to present these sum rules. Here we choose to test them in the form of relations that correspond to the symmetry under $U$-spin conjugation, see Eq.~\eqref{eq:full_flip}. Concretely, we choose to present them by testing ratios of CKM-free rates for each ``decay'' channel relative to its ``conjugate decay'' partner. The symmetry-limit expectation is
\begin{equation}\label{eq:decay_LO_RSR}
    \frac{\hat{\Gamma}(i\to f)}{\hat{\Gamma}(\overline{i} \to \overline{f})}
    = 1 + \mathcal{O}(\varepsilon)\,,
\end{equation}
where we use $\overline{i}$ and $\overline{f}$ to denote the $U$-spin conjugate initial and final states, respectively. 
Where relevant, we choose the numerator to be the decay rate with the larger error, such that the asymmetry of the resulting uncertainty is reduced.

When possible, we also test the second-order charm master sum rule, Eq.~\eqref{eq:master}. For each system, we define the corresponding ratio $R_H(\text{$U$-spin system})$ as the sum of all CF and DCS CKM-free rates divided by the sum of all SCS CKM-free rates in the system. The theoretical expectation is
\begin{equation}\label{eq:RH}
    R_H(\text{$U$-spin system}) \equiv \frac{\sum_{\text{CF,\,DCS}} \hat \Gamma}{\sum_{\text{SCS}} \hat \Gamma} =1 + \mathcal{O}(\varepsilon^2)\,.
\end{equation}

Since in the case of two-body decays the final-state kinematics is fully determined, in addition to testing the sum rules directly for rates, one can also test the sum rules for amplitudes squared. We present these tests for completeness in Appendix~\ref{app:two-body} for the relevant two-body decay systems below.

\paragraph{Second order sum rules beyond the charm master sum rule.} 

As shown in Section~\ref{sec:master}, every charm system supports at least one second-order sum rule of the form in Eq.~\eqref{eq:master}, and we report this master sum rule for all systems considered here. For systems whose external states are all $U$-spin doublets (or singlets) this master relation is the only second-order sum rule. For systems with higher irreps in the initial and/or final states, additional second-order sum rules are present. In general, some of these sum rules follow from Eqs.~\eqref{eq:sym_shmu_integer}--\eqref{eq:sym_shmu_half_integer}, while others require going beyond the simple construction presented in this work. 
Among the charm decay systems considered below, we encounter one example of an additional sum rule that follows from the construction developed in this paper and we present it explicitly for completeness (see Section~\ref{sec:neutral_charm_baryon}). We emphasize that, depending on the system, further independent second-order rate sum rules may also exist.
The rate sum rules for the examples below were cross-checked using the Mathematica package for automatic generation of $U$-spin sum
rules~\cite{KongsumrulesCode}. The package relies on the general results of Ref.~\cite{Gavrilova:2022hbx} and forthcoming work~\cite{Gavrilova:inpreparation}.

\subsection{$D^0 \rightarrow P^- P^+$}\label{sec:DtoPP}

We consider the decays of the $D^0$ meson, which is a $U$-spin singlet, into two charged pseudoscalar mesons which form $U$-spin doublets defined as:
\begin{equation}\label{eq:Ppm_def}
    P^{+} = \begin{bmatrix}
K^+ \\
\pi^+ 
\end{bmatrix} = \begin{bmatrix}
\ket{u\overline{s}} \\
-\ket{u\overline{d}}
\end{bmatrix}=\begin{bmatrix}
\ket{\frac{1}{2},+\frac{1}{2}} \\
\ket{\frac{1}{2},-\frac{1}{2}}
\end{bmatrix}, \quad P^-=\begin{bmatrix}
\pi^- \\
K^- 
\end{bmatrix} = \begin{bmatrix}
\ket{d\overline{u}} \\
\ket{s\overline{u}}
\end{bmatrix}=\begin{bmatrix}
\ket{\frac{1}{2},+\frac{1}{2}} \\
\ket{\frac{1}{2},-\frac{1}{2}}
\end{bmatrix}.
\end{equation}
This example, which we already showed in the introduction, Eqs.~(\ref{eq:DtoPP_LO_intro})-(\ref{eq:DtoPP_NLO_intro}), was studied in Ref~\cite{Grossman:2012ry} and we restate it here for completeness. The group theoretical structure of this system, Eq.~(\ref{eq:charm-structure}), is given by
\begin{equation}\label{eq:Group_DPP}
    \left(\frac{1}{2}\right)^{\otimes 2}\otimes u_H,\qquad u_H=1\;.
\end{equation}
In Table~\ref{tab:DPP} we list all decay channels of the system, together with their experimental branching ratios and CKM classification. In this example, we define the $U$-spin pairs with respect to the $P^-$ doublet in the final state: we refer to decays into $\pi^-$  as ``decays'' and decays into $K^-$ as ``conjugate decays''.
\begin{table}[t]
\centering
\resizebox{0.98\textwidth}{!}{
\begin{tabular}{|l c c | l c c|}
\hline\hline
\textbf{Decay} & \textbf{BR} & \textbf{CKM}  &
\textbf{Conjugate decay} & \textbf{BR} & \textbf{CKM} \\
\hline  & & & & & \\[-13pt]
$D^0\rightarrow \pi^{-}\pi^+$      & $(1.453\pm0.024)\cdot 10^{-3}$ & SCS &
$D^0\rightarrow K^- K^+$     & $(4.08\pm0.06)\cdot 10^{-3}$ & SCS  \\

$D^{0}\to \pi^{-} K^+$        
    & $(1.362\pm 0.025)\cdot 10^{-4}$ & DCS &
$D^{0}\to K^- \pi^+$            
    & $(3.945\pm0.030)\%$ & CF  \\

\hline\hline
\end{tabular}}
\caption{List of all the decay channels, branching ratios (BR)~\cite{PDG:2024cfk}, and CKM classification (CF/SCS/DCS) for the $D^0\rightarrow P^-P^+$ system.   }
\label{tab:DPP}
\end{table}

As we discussed in Section~\ref{sec:intro}, the symmetry-limit rate sum rules, Eq.~(\ref{eq:decay_LO_RSR}), experimentally are 
\begin{equation}
    \frac{\hat{\Gamma}(D^0 \to K^+ K^-)}{\hat{\Gamma}(D^0 \to \pi^+ \pi^-)} = 2.81 \pm 0.06\,, \qquad
    \frac{\hat{\Gamma}(D^0 \to\pi^- K^+)}{\hat{\Gamma}(D^0 \to K^-\pi^+)}
    =1.21\pm0.02 \,. \label{eq:DPP-LO}
\end{equation}
Given that the triplet in the Hamiltonian is the only higher irrep in this system, there is a single second-order rate sum rule, namely the charm master sum rule in Eq.~\eqref{eq:RH}. Using the data in Table~\ref{tab:DPP}, we find:
\begin{equation}\label{eq:master_DPP}
    R_H(D^0\rightarrow P^+P^-)=\frac{\hat{\Gamma}(D^0 \to  \pi^- K^+) + \hat{\Gamma}(D^0 \to K^- \pi^+ )}{\hat{\Gamma}(D^0 \to K^- K^+ ) + \hat{\Gamma}(D^0 \to \pi^- \pi^+ )}
    = 0.84 \pm 0.01\,. 
\end{equation} 
Evidently, the second-order rate sum rule is satisfied more accurately than the two symmetry-limit relations.

\subsection{$D^0 \rightarrow P^-V^+$ and $D^0 \rightarrow V^- P^+$}\label{sec:DtoPV}

\begin{table}[t]
\centering
\resizebox{0.98\textwidth}{!}{
\begin{tabular}{|l c c | l c c|}
\hline\hline
\textbf{Decay} & \textbf{BR} & \textbf{CKM}  &
\textbf{Conjugate decay} & \textbf{BR} & \textbf{CKM} \\
\hline & & & & & \\[-13pt]
$D^{0}\to K^- \rho^+$              
    & $(11.2 \pm 0.7)\%$ & CF  &
$D^{0}\to \pi^- K^{*+}$            
    & $3\times (1.28 \pm 0.47)\cdot 10^{-4}$ $^{\dagger}$ & DCS  \\
$D^{0}\to K^- K^{*+}$              
    & $3\times (1.55 \pm 0.10)\cdot 10^{-3}$ $(S=1.3)$ $^{\dagger}$  & SCS &
$D^{0}\to \pi^- \rho^+$            
    & $(1.01 \pm 0.05)\%$  & SCS  \\
\hline\hline  & & & & & \\[-13pt]
$D^0\rightarrow K^{*-}\pi^+$      &$3\times ( 1.82 \pm 0.30)\%$  $(S=2.2)$ $^{\dagger}$ & CF &
$D^0\rightarrow \rho^- K^+$     & $<(3.06 \pm 0.16)\cdot 10^{-4}$ $^{\ddag}$   & DCS  \\
$D^{0}\to K^{*-} K^+$        
    & $3\times (5.54\pm0.37)\cdot 10^{-4}$ $^{\dagger}$ & SCS &
$D^{0}\to \rho^- \pi^+$            
    & $(5.15 \pm 0.26)\cdot 10^{-3}$  & SCS  \\
\hline\hline
\end{tabular}}
\caption{List of decay channels, experimental branching ratios~\cite{PDG:2024cfk}, and CKM classification (CF/SCS/DCS) for the $D^0\to P^-V^+$ system (upper half) and the $D^0\to V^-P^+$ system (lower half). 
$^{\dagger}$Our average obtained from the relevant products of branching ratios involving the $K^*$ resonance given in the PDG~\cite{PDG:2024cfk}, including our symmetrization of errors. 
$^{\ddag}$Inferred bound from $\mathcal{B}(D^0\rightarrow K^+\pi^-\pi^0)$, to which $D^0\rightarrow \rho^-K^+$ contributes.
}
\label{tab:DPV}
\end{table}

We consider two-body decays of the $D^0$, a $U$-spin singlet, into a pseudoscalar and a vector meson. The rate sum rules for this system were studied in Ref.~\cite{Grossman:2012ry}, and we present them here as an example of the general charm master sum rule discussed in this paper. These decays are measured as part of Dalitz plot analyses of $D\rightarrow PPP$, see,~\emph{e.g.},~Ref.~\cite{Dery:2021mll}. Both pseudoscalars and vectors form $U$-spin doublets: the $P^\pm$ doublets are defined in Eq.~\eqref{eq:Ppm_def}, and the $V^\pm$ doublets are defined as
\begin{equation}
    V^{+} = \begin{bmatrix}
K^{*+} \\
\rho^{+} 
\end{bmatrix} = \begin{bmatrix}
\ket{u\overline{s}} \\
-\ket{u\overline{d}}
\end{bmatrix}=\begin{bmatrix}
\ket{\frac{1}{2},+\frac{1}{2}} \\
\ket{\frac{1}{2},-\frac{1}{2}}
\end{bmatrix}, \quad V^-=\begin{bmatrix}
\rho^{-} \\
K^{*-} 
\end{bmatrix} = \begin{bmatrix}
\ket{d\overline{u}} \\
-\ket{s\overline{u}}
\end{bmatrix}=\begin{bmatrix}
\ket{\frac{1}{2},+\frac{1}{2}} \\
\ket{\frac{1}{2},-\frac{1}{2}}
\end{bmatrix}.
\end{equation}
Thus, the group-theoretical structure of the $D^0 \to P^\mp V^\pm$ systems, and hence the counting and structure of the sum rules, is the same as for $D^0\to P^+P^-$. In Table~\ref{tab:DPV}, we list all decay channels for both systems \(D^0 \to P^- V^+\) and \(D^0 \to V^- P^+\), together with their experimental branching ratio and CKM classification. We define the \(U\)-spin pairs with respect to the negatively charged doublet in the final state; that is, we refer to decays into \(K^-\) or \(K^{*-}\) as ``decays'', and to decays into \(\pi^-\) or \(\rho^-\) as ``conjugate decays''.

For the \(D^0 \to P^- V^+\) system all relevant branching ratios for testing our sum rules are known, see Table~\ref{tab:DPV}. We note a subtlety regarding channels with a charged $K^*$ in the final state, see also footnote~1 in Ref.~\cite{Grossman:2012ry}.
Ref.~\cite{PDG:2024cfk} reports the product of branching ratios $\mathcal{B}(D^0\to \dots K^{*\pm})\,\mathcal{B}(K^{*\pm}\to K_S\pi^\pm)$
and $\mathcal{B}(D^0\to \dots K^{*\pm})\,\mathcal{B}(K^{*\pm}\to K^+\pi^0)$.
In order to obtain the branching ratios $\mathcal{B}(D^0\to \dots K^{*\pm})$ from that, we use the isospin relation~\cite{Grossman:2012ry}
\begin{align}
\mathcal{B}(K^{*\pm}\rightarrow K^{\pm}\pi^0)=\frac{1}{3}\,,\\
\mathcal{B}(K^{*\pm}\rightarrow K^0\pi^{\pm})=\frac{2}{3}\,,
\end{align}
implying $\mathcal{B}(K^{*\pm}\rightarrow K_S\pi^{\pm})=1/3$.
If both channels are measured, we use a weighted average of the two. 
We note that in the case of the branching ratios 
$\mathcal{B}(D^0\rightarrow K^-K^{*+})$ and $\mathcal{B}(D^0\rightarrow K^{*-}\pi^+)$ the two averaged inputs are in slight tension with each other.
In order to account for that, following the PDG prescription~\cite{PDG:2024cfk} in these cases we increase the error by a scale factor $S=\sqrt{\chi^2/(N-1)}$ with $N=2$.
To be specific we specify the relevant applied scale factors explicitly in Table~\ref{tab:DPV}. Future experimental work to further clarify the situation would be very beneficial in order to improve the quality of our sum rule tests following below.

We numerically test the symmetry-limit rate sum rules in Eq.~\eqref{eq:decay_LO_RSR}:
\begin{align}
   \frac{
    \hat{\Gamma}(D^0 \to \pi^- K^{*+})
    }{
     \hat{\Gamma}(D^0 \to K^- \rho^+)
    }
    &= 1.21\pm0.45\;,  \,\label{eq:RLO-DPV-1}\\
    \frac{\hat{\Gamma}(D^0 \to K^- K^{*+})}{\hat{\Gamma}(D^0 \to \pi^- \rho^+)} &= 0.46 \pm 0.04\;. \label{eq:RLO-DPV-2}
\end{align}
For the second order rate sum rule, Eq.~(\ref{eq:RH}), we find 
\begin{equation}\label{eq:RH-DPV}
    R_H(D^0 \to P^- V^+) =
    \frac{\hat{\Gamma}(D^0 \to K^- \rho^+) + \hat{\Gamma}(D^0 \to \pi^- K^{*+})}
         {\hat{\Gamma}(D^0 \to K^- K^{*+}) + \hat{\Gamma}(D^0 \to \pi^- \rho^+)}
    =  0.89\pm 0.18\,.
\end{equation}
Given the large experimental uncertainties, it is difficult to draw a conclusive statement about the $U$-spin breaking expansion. Still, we note that the central value of \(R_H(D^0\to P^-V^+)\) is closer to unity than the ratio in Eq.~\eqref{eq:RLO-DPV-2}, while it is comparable, within uncertainties, to the ratio in Eq.~\eqref{eq:RLO-DPV-1}. The data is therefore compatible with the assumption that $\varepsilon$ is ``small''.

For the \(D^0 \to V^- P^+\) system, all branching ratios are known except for \(D^0 \to \rho^- K^+\). Using the fact that the branching ratio of the three-body decay is bigger than any of the contributing pseudo two-body decays, we demand that $\mathcal{B}(D^0\rightarrow \rho^- K^+) < \mathcal{B}(D^0\rightarrow K^+\pi^-\pi^0)$, see Table \ref{tab:DPV}. For this system, only one symmetry-limit rate sum rule can be tested:
\begin{equation}
     \frac{\hat{\Gamma}(D^0 \to K^{*-} K^+)}{\hat{\Gamma}(D^0 \to \rho^- \pi^+)} = 0.32 \pm 0.03\;, \label{eq:DVP-LO}
\end{equation}
and, instead of testing the second-order master sum rule in Eq.~(\ref{eq:RH}), we use it to predict the unmeasured branching fraction. To do this, we rewrite the master sum rule as
\begin{equation}
    \hat{\Gamma}(D^0\rightarrow \rho^-K^{+}) =
    -\hat{\Gamma}(D^0\rightarrow K^{*-}\pi^{+})+ \hat{\Gamma}(D^0\rightarrow K^{*-}K^{+}) + \hat{\Gamma}(D^0\rightarrow \rho^{-}\pi^{+})+\mathcal{O}(\varepsilon^2).
\end{equation}
Using the inputs in Table~\ref{tab:DPV}, this yields the prediction
\begin{equation}\label{prediction:DPV}
    \mathcal{B}(D^0\rightarrow \rho^-K^{+})=\left(2.08 \pm 0.30\right)\cdot 10^{-4}\,,
\end{equation}
which is consistent with the current experimental bound listed in Table~\ref{tab:DPV}.
Here and in the following, in our predictions we only quote the experimental error. A generic overall theory error of $\mathcal{O}(\varepsilon^2)$ is implied throughout.

\subsection{$D_{q}^- \rightarrow P^+ P^- P^-$}\label{sec:DtoPPP}

\begin{table}[t]
\centering
\resizebox{0.98\textwidth}{!}{
\begin{tabular}{|l c c | l c c|}
\hline\hline
\textbf{Decay} & \textbf{BR} & \textbf{CKM}  &
\textbf{Conjugate decay} & \textbf{BR} & \textbf{CKM} \\
\hline & & &  & & \\[-13pt]
$D^-\rightarrow\pi^+K^-\pi^-$      & $(4.91\pm0.09)\cdot10^{-4}$ & DCS &
$D^-_s\rightarrow K^+\pi^-K^-$     & $(5.45\pm0.08)\cdot10^{-2}$  & CF  \\

$D^-\rightarrow K^+K^-K^-$         & $(6.14\pm0.11)\cdot10^{-5}$ & DCS &
$D^-_s\rightarrow \pi^+\pi^-\pi^-$ & $(1.090\pm0.014)\cdot10^{-2}$   & CF  \\

$D^-\rightarrow K^+\pi^-\pi^-$     & $(9.38\pm0.16)\cdot10^{-2}$ & CF  &
$D^-_s\rightarrow \pi^+K^-K^-$     & $(1.293\pm0.027)\cdot10^{-4}$  & DCS \\

$D^-\rightarrow K^+K^-\pi^-$       & $(9.68\pm0.18)\cdot10^{-3}$ & SCS &
$D^-_s\rightarrow \pi^+\pi^-K^-$   & $(6.23\pm0.10)\cdot10^{-3}$  & SCS \\

$D^-\rightarrow \pi^+\pi^-\pi^-$   & $(3.27\pm0.09)\cdot10^{-3}$ & SCS &
$D^-_s\rightarrow K^+K^-K^-$       & $(2.18\pm0.20)\cdot10^{-4}$  & SCS \\[1pt]
\hline\hline
\end{tabular}}
\caption{List of decay channels, branching ratios (BR)~\cite{PDG:2024cfk}, and CKM classification (CF/SCS/DCS) for the $D^-_{q}\to P^+P^-P^-$ system. Within each pair, the ``decay'' and ``conjugate decay'' are defined relative to $D_q^-$.}
\label{tab:DPPP}
\end{table}

We consider the system of three-body charged \(D\)-meson decays to pseudoscalar mesons,
\(D_q^-\to P^+ P^- P^-\).
In this system, all external particles transform as components of \(U\)-spin doublets.
The \(P^\pm\) doublets are defined in Eq.~\eqref{eq:Ppm_def}, and the \(D_q^-\) doublet is given by
\begin{equation}\label{eq:D_ds_def}
D_q^-\equiv
\begin{bmatrix}
D^-\\
D_s^-
\end{bmatrix}
=
\begin{bmatrix}
\ket{d\overline{c}}\\
\ket{s\overline{c}}
\end{bmatrix}
=
\begin{bmatrix}
\ket{\frac{1}{2},+\frac{1}{2}}\\
\ket{\frac{1}{2},-\frac{1}{2}}
\end{bmatrix}\,.
\end{equation}
Thus the group-theoretic structure for this system is given by the following tensor product:
\begin{equation}
    \left(\frac{1}{2}\right)^{\otimes 4}\otimes u_H,\quad u_H=1\;.
\end{equation}
The full list of decays in this system is given in Table~\ref{tab:DPPP}, including the measured branching ratios and the CKM classification for each decay. We use the initial-state doublet $D_q^-$ to organize the decays into pairs. We refer to $D^-$ decays as ``decays'' and to $D_s^-$ decays as ``conjugate decays''. As the final state contains two identical multiplets $P^-$, this system also provides a concrete example of the identical-particle treatment developed in Section~\ref{sec:identical}. We work out this example explicitly in Appendix~\ref{app:identical_DPPP}.

Since all branching ratios are measured, we can assess the quality of the $U$-spin--breaking expansion by confronting the charm master sum rule in Eq.~\eqref{eq:master_ratio_second_order} with data, together with the symmetry-limit relations in Eq.~\eqref{eq:decay_LO_RSR}. Using the data in Table~\ref{tab:DPPP}, we find:
\begin{align}
    \frac{\hat{\Gamma}(D^- \rightarrow \pi^+ K^- \pi^-)}{\hat{\Gamma}(D^-_s \rightarrow K^+ K^- \pi^-)}&= 1.537\pm 0.036\,, \label{eq:DPPP-LO-1}\\
    \frac{\hat{\Gamma}(D^- \rightarrow K^+ K^- K^-)}{\hat{\Gamma}(D^-_s \rightarrow \pi^+ \pi^- \pi^-)} &=0.961\pm0.021\,,\label{eq:DPPP-LO-2} \\
    \frac{\hat{\Gamma}(D^- \rightarrow K^+ \pi^- \pi^-)}{\hat{\Gamma}(D^-_s \rightarrow \pi^+ K^- K^-)} &=1.001\pm0.027\,, \label{eq:DPPP-LO-3}\\
    \frac{\hat{\Gamma}(D^- \rightarrow K^+ K^- \pi^-)}{\hat{\Gamma}(D^-_s \rightarrow \pi^+ \pi^- K^-)}&= 0.754\pm 0.019\,, \label{eq:DPPP-LO-4}\\
    \frac{
        \hat{\Gamma}(D^-_s \rightarrow K^+ K^- K^-)
    }{
         \hat{\Gamma}(D^- \rightarrow \pi^+ \pi^- \pi^-)
        } &= 0.138\pm0.013\,. \label{eq:DPPP_LO_data_large}
\end{align}
As in the two-body decays considered in Sections~\ref{sec:DtoPP}--\ref{sec:DtoPV}, the charm Hamiltonian is the only higher irrep in this system. Thus, there is only one second-order rate sum rule, given in Eq.~\eqref{eq:RH}. To present this sum rule explicitly, we define the sum of all CF and DCS CKM-free decay rates of the system as
\begin{multline}
    \sum\nolimits_{\text{CF, DCS}}\hat{\Gamma}
    = \hat{\Gamma}(D_s^- \rightarrow \pi^+ \pi^- \pi^-)
    + \hat{\Gamma}(D^- \rightarrow K^+ K^- K^-)
    + \hat{\Gamma}(D_s^- \rightarrow \pi^+ K^- K^-)\\
    + \hat{\Gamma}(D^- \rightarrow K^+ \pi^- \pi^-)
    + \hat{\Gamma}(D_s^- \rightarrow K^+ \pi^- K^-)
    + \hat{\Gamma}(D^- \rightarrow \pi^+ K^- \pi^-)\;,
\end{multline}
and the sum of all SCS CKM-free decay rates as
\begin{align}
    \sum\nolimits_{\text{SCS}}\hat{\Gamma}
    = \hat{\Gamma}(D_s^- \rightarrow \pi^+ \pi^- K^-)
    + \hat{\Gamma}(D^- \rightarrow K^+ K^- \pi^-) \\
    + \hat{\Gamma}(D_s^- \rightarrow K^+ K^- K^-)
    + \hat{\Gamma}(D^- \rightarrow \pi^+ \pi^- \pi^-)\,. \nonumber
\end{align}
Using the data in Tables~\ref{tab:numerical} and~\ref{tab:DPPP}, we check the second order sum rule numerically:
\begin{equation}\label{eq:RH_DPPP_data}
    R_H(D_q^- \rightarrow P^+P^-P^-)=\frac{\sum_{\text{CF, DCS}}\hat{\Gamma}}{\sum_{\text{SCS}}\hat{\Gamma}}
    =1.050\pm 0.015\,.
\end{equation}

Looking at the data we see that the central values of the LO ratios vary. In particular, Eq.~\eqref{eq:DPPP_LO_data_large} shows a sizable departure from the symmetry limit. Yet, the second-order ratio in Eq.~\eqref{eq:RH_DPPP_data} remains very close to $1$. Thus, we conclude that the data on $D_q^-$ decays supports the hypothesis of ``small'' $\varepsilon$ and $U$-spin being a good approximate symmetry.

\subsection{Charged charm baryon decays: $C_b^+\to L_b^+P^+P^-$ \label{sec:charged_charm_baryon} }

\begin{table}[t]
\centering
\resizebox{0.98\textwidth}{!}{%
\begin{tabular}{|l c c | l c c|}
\hline\hline
\textbf{Decay} & \textbf{BR} & \textbf{CKM} &
\textbf{Conjugate decay} & \textbf{BR} & \textbf{CKM} \\
\hline
& & & & & \\[-13pt]
$\Lambda_c^+ \to \Sigma^+ K^- K^+$     & $(3.6\pm0.4)\cdot10^{-3}$  & CF  &
$\Xi_c^+ \to p \pi^- \pi^+$            & ?                           & DCS \\

$\Lambda_c^+ \to \Sigma^+ \pi^- \pi^+$ & $(4.54\pm0.20)\%$           & CF  &
$\Xi_c^+ \to p K^- K^+$                & ?                           & DCS \\

$\Lambda_c^+ \to \Sigma^+ \pi^- K^+$   & $(2.04\pm0.26)\cdot10^{-3}$  & SCS &
$\Xi_c^+ \to p K^- \pi^+$              & $(6.2\pm3.0)\cdot10^{-3}$   & SCS \\

$\Lambda_c^+ \to p K^- \pi^+$          & $(6.35\pm0.25)\%$           & CF  &
$\Xi_c^+ \to \Sigma^+ \pi^- K^+$       & ?                           & DCS \\

$\Lambda_c^+ \to p K^- K^+$            & $(1.08\pm0.05)\cdot10^{-3}$  & SCS &
$\Xi_c^+ \to \Sigma^+ \pi^- \pi^+$     & $(1.4\pm0.8)\%$             & SCS \\

$\Lambda_c^+ \to p \pi^- \pi^+$        & $(4.67\pm0.24)\cdot10^{-3}$  & SCS &
$\Xi_c^+ \to \Sigma^+ K^- K^+$         & $(4.3\pm2.5)\cdot10^{-3}$   & SCS \\

$\Lambda_c^+ \to p \pi^- K^+$          & $(1.13\pm0.17)\cdot10^{-4}$  & DCS &
$\Xi_c^+ \to \Sigma^+ K^- \pi^+$       & $(2.7\pm1.2)\%$             & CF  \\[2pt]
\hline\hline
\end{tabular}}
\caption{List of all decay channels, with respective $U$-spin conjugate, branching ratios (BR)~\cite{PDG:2024cfk}, and CKM classification (CF/SCS/DCS) for the $C_b^+\rightarrow L_b^+P^+P^-$ system.}
\label{tab:Cb+}
\end{table}

We consider the decay systems of charged charm baryons, which transform as a \(U\)-spin doublet, defined as
\begin{equation}\label{eq:Cb+}
    C_b^+ = \begin{bmatrix}
\Lambda_c^+ \\
\Xi_c^+ 
\end{bmatrix} = \begin{bmatrix}
\ket{udc} \\
\ket{usc}
\end{bmatrix}=\begin{bmatrix}
\ket{\frac{1}{2},+\frac{1}{2}} \\
\ket{\frac{1}{2},-\frac{1}{2}}
\end{bmatrix}.
\end{equation}
In particular, we consider the three body decays of charged charm baryons into one positively charged light baryon and two pseudoscalar mesons. The $P^{\pm}$ doublets are defined in Eq. (\ref{eq:Ppm_def}), and the positively charged light baryons form a $U$-spin doublet given by
\begin{equation}\label{eq:Lb+}
L_b^+= \begin{bmatrix}
p \\
\Sigma^+ 
\end{bmatrix} = \begin{bmatrix}
\ket{uud} \\
\ket{uus}
\end{bmatrix}=\begin{bmatrix}
\ket{\frac{1}{2},+\frac{1}{2}} \\
\ket{\frac{1}{2},-\frac{1}{2}}
\end{bmatrix}.
\end{equation}
Hence, the group theoretical structure of this system is
\begin{equation}\label{eq:group_Cb+}
    \left(\frac{1}{2}\right)^{\otimes 4}\otimes u_H,\quad u_H=1\;.
\end{equation}
In Table~\ref{tab:Cb+}, we list all decay channels in this \(U\)-spin system, together with the experimental branching ratios for the processes that have been measured and their CKM classification. In this case, we use the initial-state doublet \(C_b^+\) to define the \(U\)-spin pairs: we refer to \(\Lambda_c^+\) decays as ``decays'' and to \(\Xi_c^+\) decays as ``conjugate decays''.  
Since not all decay rates in this system have been measured yet, we can numerically test only those symmetry-limit rate sum rules, Eq.~(\ref{eq:decay_LO_RSR}), for which data is available:
\begin{align}
    \frac{
        \hat{\Gamma}(\Xi_c^+\rightarrow p K^-\pi^+) 
    }{
        \hat{\Gamma}(\Lambda_c^+\rightarrow \Sigma^+ \pi^- K^+)
    } &=  1.38\pm 0.69\,, \\
    \frac{
        \hat{\Gamma}(\Xi_c^+\rightarrow \Sigma^+ \pi^-\pi^+)  
    }{
        \hat{\Gamma}(\Lambda_c^+\rightarrow p K^- K^+)  
    } &=  5.81\pm 3.33\,, \\
    \frac{
        \hat{\Gamma}(\Xi_c^+\rightarrow \Sigma^+ K^-K^+)
    }{
        \hat{\Gamma}(\Lambda_c^+\rightarrow p \pi^- \pi^+) 
    } &= 0.41\pm0.24\,,\\
    \frac{
        \hat{\Gamma}(\Xi_c^+\rightarrow \Sigma^+ K^-\pi^+)
        }{
        \hat{\Gamma}(\Lambda_c^+\rightarrow p \pi^- K^+)  
    } &=0.30\pm0.14\,.
\end{align}
In addition, we explicitly write the second order master rate sum rule, Eq. (\ref{eq:RH}), for this system:
\begin{align}
    &\hat{\Gamma}(\Lambda_c^+\rightarrow\Sigma^+ K^- K^+) + \hat{\Gamma}(\Xi_c^+\rightarrow p \pi^- \pi^+) + \hat{\Gamma}(\Lambda_c^+\rightarrow\Sigma^+ \pi^- \pi^+) 
    + \hat{\Gamma}(\Xi_c^+\rightarrow p K^- K^+) \nonumber\\ 
    +\; & \hat{\Gamma}(\Lambda_c^+\rightarrow p K^- \pi^+) + \hat{\Gamma}(\Xi_c^+\rightarrow\Sigma^+ \pi^- K^+) 
    + \hat{\Gamma}(\Lambda_c^+\rightarrow p \pi^- K^+) + \hat{\Gamma}(\Xi_c^+\rightarrow\Sigma^+ K^- \pi^+) \nonumber \\
    =\; & \hat{\Gamma}(\Lambda_c^+\rightarrow\Sigma^+ \pi^- K^+) + \hat{\Gamma}(\Xi_c^+\rightarrow p K^- \pi^+) + \hat{\Gamma}(\Lambda_c^+\rightarrow p K^- K^+)  + \hat{\Gamma}(\Xi_c^+\rightarrow\Sigma^+ \pi^- \pi^+) \nonumber\\  
    +\; & \hat{\Gamma}(\Lambda_c^+\rightarrow p \pi^- \pi^+) +\, \hat{\Gamma}(\Xi_c^+\rightarrow\Sigma^+ K^- K^+)+\mathcal{O}(\varepsilon^2)\;.
\end{align}
As shown in Table~\ref{tab:Cb+}, there are not enough measurements to perform a numerical test of the second-order sum rule and to compute \(R_H(C_b^+ \to L_b^+ P^+ P^-)\). Nevertheless, the sum rule can be used to provide a theoretical prediction for the sum of the branching ratios of the decay modes that have not yet been measured,
\begin{equation}\label{prediction:charged_charm}
    \mathcal{B}(\Xi_c^+ \to p \pi^- \pi^+)
    + \mathcal{B}(\Xi_c^+ \to p K^- K^+)
    + \mathcal{B}(\Xi_c^+ \to \Sigma^+ \pi^- K^+)
    = \left( 1.19 \pm 0.48 \right) \cdot 10^{-3}\,.
\end{equation}
This prediction is consistent, at the order-of-magnitude level, with the expected size of the sum of the three DCS branching fractions.

\subsection{Neutral charm baryon decays:  $\Xi_c^0\rightarrow L_b^-P^+$, $\Xi_c^0\rightarrow L_b^+ P^-$, and $\Xi_c^0\to\Omegab^-P^+$ \label{sec:neutral_charm_baryon}}

\begin{table}[t]
\centering
\resizebox{0.98\textwidth}{!}{
\begin{tabular}{|l c c | l c c|}
\hline\hline
\textbf{Decay} & \textbf{BR} & \textbf{CKM}  &
\textbf{Conjugate decay} & \textbf{BR} & \textbf{CKM} \\
\hline & & & & & \\[-13pt]
$\Xi_c^0\rightarrow \Xi^-\pi^+$             
    & $(1.43 \pm 0.27)\%$  & CF  &
$\Xi_c^0\rightarrow \Sigma^- K^+$            
    & ?  & DCS  \\
$\Xi_c^0\rightarrow \Sigma^-\pi^+$            
    & ? & SCS & $\Xi_c^0\rightarrow \Xi^-K^+$              
    & $(3.9 \pm 1.1)\cdot 10^{-4}$  & SCS  \\[2pt]
\hline \hline  & & & & & \\[-13pt]
$\Xi_c^0\rightarrow p K^-$      & ?   & SCS &
$\Xi_c^0\rightarrow \Sigma^+\pi^-$     & ? & SCS  \\
$\Xi_c^0\rightarrow \Sigma^+ K^-$            
    & $(1.8\pm0.4)\cdot 10^{-3}$ & CF  & $\Xi_c^0\rightarrow p\pi^-$       
    & ?  & DCS \\[2pt]
\hline \hline & & & & & \\[-13pt]
$\Xi_c^0\rightarrow\Delta^-\pi^+$     & ?      & DCS &
$\Xi_c^0 \to \Omega^{-} K^+$           & ?     & CF  \\

$\Xi_c^0\rightarrow\Sigma^{*-}\pi^+$  & ?      & SCS &
$\Xi_c^0 \to \Xi^{*-} K^+$           & ?       & SCS \\

$\Xi_c^0\rightarrow\Xi^{*-}\pi^+$    & ?       & CF  &
$\Xi_c^0 \to \Sigma^{*-} K^+$     & ?          & DCS \\[2pt]
\hline\hline
\end{tabular}}
\caption{List of decay channels, branching ratios (BR)~\cite{PDG:2024cfk}, and CKM classification (CF/SCS/DCS) for the $\Xi_c^0\to L_b^-P^+$ system (upper half), the $\Xi_c^0\to L_b^+P^-$ system (central part), and the $\Xi_c^0\rightarrow \Omegab^-P^+$ system (lower half).}
\label{tab:Xi0_LbP}
\end{table}

We consider the systems of decays of the neutral charm baryon $\Xi_c^0$, which is to a very good approximation a $U$-spin singlet~(see Table~\ref{tab:numerical})
\begin{equation}
\ket{\Xi_c^0}=\ket{csd}=\ket{0,0}\,.
\end{equation}
Here we neglect  $\Xi_c^0- \Xi_c^{'0}$ mixing, which is at the order of $1\%$~\cite{Brown:2014ena, Liu:2023feb, Sun:2023noo, Liu:2023pwr}  and thus very small.
We consider three decay systems of $\Xi_c^0$ into a charged baryon and a pseudoscalar meson. The pseudoscalar mesons doublet is defined in Eq.~(\ref{eq:Ppm_def}). For the charged baryon in the final state, we consider the positively charged light-baryon multiplet $L_b^+$, the negatively charged light-baryon multiplet $L_b^-$, and the $\Omegab^-$ baryon multiplet. The $L_b^+$ doublet is defined in Eq. (\ref{eq:Lb+}), and the transformation properties of $L_b^-$ and $\Omegab^-$ are given by
\begin{equation}\label{eq:Lb-}
L_b^-=
\begin{bmatrix}
\Sigma^-    \\
\Xi^-
\end{bmatrix} 
= \begin{bmatrix}
\ket{sdd} \\
\ket{ssd}
\end{bmatrix}=\begin{bmatrix}
\ket{\frac{1}{2},+\frac{1}{2}} \\
\ket{\frac{1}{2},-\frac{1}{2}}
\end{bmatrix}\;, \qquad\Omegab^- = \begin{bmatrix}
\Delta^- \\
\Sigma^{*-} \\
\Xi^{*-} \\
\Omega^-
\end{bmatrix} = \begin{bmatrix}
\ket{ddd} \\
\ket{dds} \\
\ket{dss} \\
\ket{sss}
\end{bmatrix}=\begin{bmatrix}
\ket{\frac{3}{2},+\frac{3}{2}} \\
\ket{\frac{3}{2},+\frac{1}{2}} \\
\ket{\frac{3}{2},-\frac{1}{2}} \\
\ket{\frac{3}{2},-\frac{3}{2}}
\end{bmatrix}.
\end{equation}
Thus, we consider the three different systems: $\Xi_c^0\rightarrow L_b^-P^+$, $\Xi_c^0\rightarrow L_b^+ P^-$, and $\Xi_c^0\to\Omegab^-P^+$. In Table~\ref{tab:Xi0_LbP}, we list all decay channels for these systems, together with the experimental branching ratios (when available) and their CKM classification. In this example, we define the \(U\)-spin pairs with respect to the \(P^{\pm}\) doublet in the final state: we refer to decay modes with \(\pi^+\) or $K^-$ in the final state as ``decays'' and to those with a \(K^+\) or $\pi^-$ in final state as ``conjugate decays''.

The systems $\Xi_c^0\rightarrow L_b^-P^+$ and $\Xi_c^0\rightarrow L_b^+ P^-$ have the same group theoretical structure:
\begin{equation}
    \left(\frac{1}{2}\right)^{\otimes 2}\otimes u_H\;,\qquad u_H=1\;.
\end{equation}
  For both the systems, not all decay rates have been measured, and the available data are insufficient to test either the trivial leading-order rate sum rule in Eq.~(\ref{eq:decay_LO_RSR}) or the second-order rate sum rule in Eq.~(\ref{eq:RH}), which we report here explicitly for the $\Xi_c^0\rightarrow L_b^-P^+$ system: 
\begin{align}
    \frac{\hat{\Gamma}\left(\Xi_c^0\rightarrow \Xi^- \pi^+\right) + \hat{\Gamma}\left(\Xi_c^0\rightarrow \Sigma^- K^+\right)}{
    \hat{\Gamma}\left(\Xi_c^0\rightarrow \Xi^- K^+\right) + \hat{\Gamma}\left(\Xi_c^0\rightarrow \Sigma^- \pi^+\right)
    } =  1 + \mathcal{O}(\varepsilon^2)\;,
\end{align}
and for the $\Xi_c^0\rightarrow L_b^+ P^-$ system:
\begin{equation}
    \frac{
       \hat{\Gamma}\left(\Xi_c^0\rightarrow p \pi^-\right) + \hat{\Gamma}\left(\Xi_c^0\rightarrow \Sigma^+ K^-\right)    
        }{
     \hat{\Gamma}\left(\Xi_c^0\rightarrow p K^-\right) + \hat{\Gamma}\left(\Xi_c^0\rightarrow \Sigma^+ \pi^-\right)
    }= 1 + \mathcal{O}(\varepsilon^2)\;.
\end{equation}
We solve the above relations for the unmeasured branching ratios, giving
\begin{align}
\frac{\mathcal{B}(\Xi_c^0\rightarrow \Sigma^-\pi^+)}{f_{\mathrm{SCS}}} - 
\frac{\mathcal{B}(\Xi_c^0\rightarrow \Sigma^- K^+ )}{f_{\mathrm{DCS}}} &= 
\frac{\mathcal{B}(\Xi_c^0\rightarrow \Xi^-\pi^+)}{ f_{\mathrm{CF}}} - 
\frac{\mathcal{B}(\Xi_c^0\rightarrow \Xi^- K^+ )}{f_{\mathrm{SCS}}}\,,
\end{align}
and
\begin{align}
\frac{\mathcal{B}\left(\Xi_c^0\rightarrow p K^-\right)}{f_{SCS}} + 
\frac{\mathcal{B}\left(\Xi_c^0\rightarrow \Sigma^+ \pi^-\right)}{f_{SCS}} -
\frac{\mathcal{B}\left(\Xi_c^0\rightarrow p \pi^-\right)}{f_{DCS}} &=  \frac{\mathcal{B}\left(\Xi_c^0\rightarrow \Sigma^+ K^-\right)}{f_{CF}}\,.
\end{align}
Numerically, with current data we obtain the predictions
\begin{align}\label{prediction:neutral_charm-}
\frac{\mathcal{B}(\Xi_c^0\rightarrow \Sigma^-\pi^+)}{f_{SCS}} - 
\frac{\mathcal{B}(\Xi_c^0\rightarrow \Sigma^- K^+ )}{f_{DCS}} &= (7.75 \pm 3.77)\cdot 10^{-3}  \,, \\
\label{prediction:neutral_charm+}
\frac{\mathcal{B}\left(\Xi_c^0\rightarrow p K^-\right)}{f_{SCS}} + \frac{\mathcal{B}\left(\Xi_c^0\rightarrow \Sigma^+ \pi^-\right)}{f_{SCS}} -
\frac{\mathcal{B}\left(\Xi_c^0\rightarrow p \pi^-\right)}{f_{DCS}} &= (2.00\pm 0.44)\cdot 10^{-3}\,.
\end{align}
On the other hand, the system $\Xi_c^0\to\Omegab^-P^+$ has the group theoretical structure:
\begin{equation}\label{eq:structure:Cb0_Omega}
    \frac{1}{2}\otimes\frac{3}{2}\otimes u_H,\qquad u_H=1\;.
\end{equation}
Also in this case, since no branching ratios have been measured for this system, we can test neither the symmetry-limit sum rules nor the second-order master rate sum rule, which we report here:
\begin{align}
\hat{\Gamma}\!\left(\Xi_c^0\to \Delta^-\pi^+\right)
+\hat{\Gamma}\!\left(\Xi_c^0\to \Omega^- K^+\right)
+\hat{\Gamma}\!\left(\Xi_c^0\to \Xi^{*-}\pi^+\right)
+\hat{\Gamma}\!\left(\Xi_c^0\to \Sigma^{*-} K^+\right)
\nonumber\\
=\hat{\Gamma}\!\left(\Xi_c^0\to \Sigma^{*-}\pi^+\right)
+\hat{\Gamma}\!\left(\Xi_c^0\to \Xi^{*-} K^+\right) + \mathcal{O}(\varepsilon^2)\;.
\end{align}

For this system, the construction developed in this paper gives an additional second-order rate sum rule. This sum rule corresponds to the Shmushkevich equation written for the multiplet $\Omegab^-$, i.e.\ Eq.~\eqref{eq:sym_shmu_half_integer} with $I_i=3/2$. In analogy with Eq.~\eqref{eq:Gamma_inclusive_H}, we denote the inclusive CKM-free rate for a fixed component of irrep $I_i$ with $m$-QN $m_i$ by $\hat{\Gamma}_{i}(m_{i})$. With this notation, Eq.~\eqref{eq:sym_shmu_half_integer} yields the following general second-order rate sum rule in terms of inclusive rates:
\begin{equation}\label{app:eq:RSR_quadruplet}
    I_i = \frac{3}{2}:\qquad \hat{\Gamma}_{i}\left(+\frac{1}{2}\right) + \hat{\Gamma}_{i}\left(-\frac{1}{2}\right) = \hat{\Gamma}_{i}\left(+\frac{3}{2}\right) + \hat{\Gamma}_{i}\left(-\frac{3}{2}\right) + \mathcal{O}(\varepsilon^2)\;.
\end{equation}
In the case of $\Xi_c^0\to \Omegab^- P^+$ decays, this relation implies that, up to second order in $U$-spin breaking, the sum of CKM-free decay rates with $\Sigma^{*-}$ and $\Xi^{*-}$ in the final state is equal to the sum of CKM-free decay rates with $\Delta^-$ and $\Omega^-$ in the final state. Explicitly, we find:
\begin{multline}\label{app:eq:additional_Xi0_Omega}
\hat{\Gamma}\left(\Xi_c^0\to\Sigma^{*-}K^+\right) + 
\frac{1}{2}\hat{\Gamma}\left(\Xi_c^0\to\Sigma^{*-}\pi^+\right) + 
\hat{\Gamma}\left(\Xi_c^0\to\Xi^{*-}\pi^+\right) +
\frac{1}{2} \hat{\Gamma}\left(\Xi_c^0\to\Xi^{*-}K^+\right)  \\ = 
\hat{\Gamma}\left(\Xi_c^0\to\Delta^-\pi^+\right) + \hat{\Gamma}\left(\Xi_c^0\to\Omega^-K^+\right) +\mathcal{O}(\varepsilon^2)\;,
\end{multline}
where the factors of $1/2$ multiplying the SCS rates follow from our convention for CKM-free rates, see Eqs.~\eqref{eq:f_CKM} and
\eqref{eq:Gammahat_from_data}.

\section{Conclusions and outlook}\label{sec:conclusions}

\begin{figure}[t]
    \centering
    \includegraphics[width=0.7\textwidth]{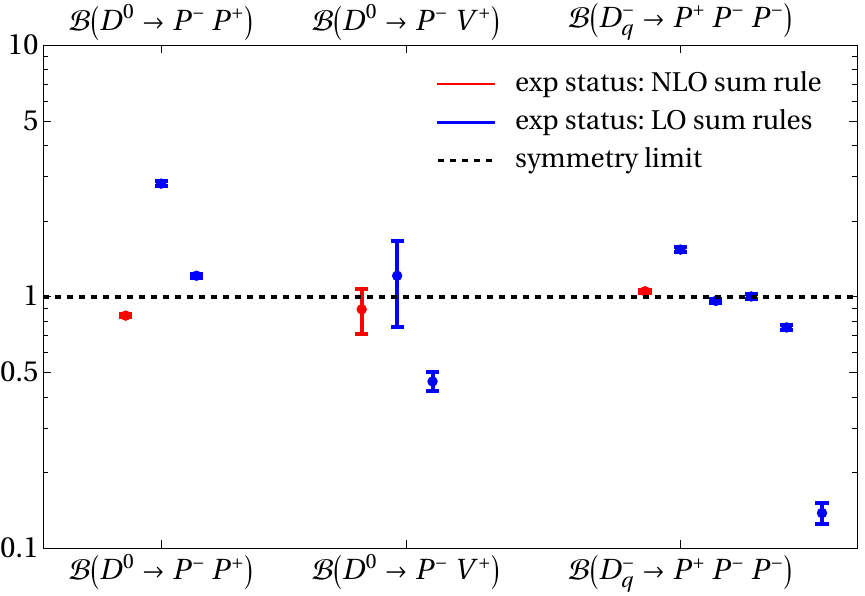}
    \caption{%
        Current experimental determinations of the NLO sum rules (red) Eqs.~(\ref{eq:master_DPP}, \ref{eq:RH-DPV}, \ref{eq:RH_DPPP_data}) in comparison to the corresponding LO sum rules (blue) Eqs.~(\ref{eq:DPP-LO}, \ref{eq:RLO-DPV-1}, \ref{eq:RLO-DPV-2}, \ref{eq:DPPP-LO-1}, \ref{eq:DPPP-LO-2}, \ref{eq:DPPP-LO-3}, \ref{eq:DPPP-LO-4}). The $U$-spin limit is illustrated by the dashed black line.}
    \label{fig:plot}
\end{figure}

\paragraph{Summary of the results.} There are two main points in this paper. The first is theoretical: we present a generic method for deriving $SU(2)_F$ rate sum rules that are valid up to second order in the symmetry breaking. We show that any leading-order sum rule that is invariant under the exchange of the two flavors related by $SU(2)_F$ remains valid through second order. We then use the Shmushkevich method to identify, for a given set of irreps in the system, the symmetry-limit sum rules that are symmetric under this flavor exchange and therefore hold to second order. Our main theoretical results are the master sum rules in Eqs.~\eqref{eq:sym_shmu_integer}--\eqref{eq:sym_shmu_half_integer}.

The second main point is phenomenological. We apply our general group-theoretic result to weak charm decays and show that, neglecting $\mathcal{O}(\lambda^4)\sim 10^{-3}$ effects of the third generation, it implies a master sum rule in Eq.~\eqref{eq:master}. This second-order sum rule is universal for all systems of hadronic weak charm decays. We then confront this prediction with available data. When analyzing the data we make several approximations as outlined in Section~\ref{sec:examples}.
For the systems where all decay channels are measured the current experimental status is summarized in Fig.~\ref{fig:plot}. In all three accessible examples, the second-order sum rule is satisfied to high accuracy. This is in contrast to the first-order sum rules, for which the experimental outcomes vary widely, ranging from good agreement to large deviations from the symmetry limit. In cases where not all branching ratios entering a second-order sum rule are measured, we use the available data to predict individual branching ratios or linear combinations of branching ratios. These predictions hold to second order in $U$-spin breaking and are summarized in Table~\ref{tab:NLO_predictions-sum}.

\begin{table}[t]
    \centering
    \renewcommand{\arraystretch}{1.4}
    \begin{tabular}{|l|c|}
\hline
\hline
Predicted quantity & Numerical value \\
        \hline
        $\mathcal{B}(D^0\rightarrow \rho^- K^+)$
        & $(2.08\pm 0.30)\cdot 10^{-4}$ \\
        \hline
        $\mathcal{B}(\Xi_c^+ \to p \pi^- \pi^+)
        + \mathcal{B}(\Xi_c^+ \to p K^- K^+)
        + \mathcal{B}(\Xi_c^+ \to \Sigma^+ \pi^- K^+)$
        & $(1.19 \pm 0.48)\cdot 10^{-3}$ \\
        \hline \\[-38pt] & \\
        $\dfrac{\mathcal{B}(\Xi_c^0\rightarrow \Sigma^-\pi^+)}{f_{SCS}}
        - \dfrac{\mathcal{B}(\Xi_c^0\rightarrow \Sigma^- K^+ )}{f_{DCS}}$
        & $(7.75\pm 3.77)\cdot 10^{-3}$ \\[-16pt] & \\
        \hline \\[-36pt] & \\
        $\dfrac{\mathcal{B}(\Xi_c^0\rightarrow p K^-)}{f_{SCS}}
        + \dfrac{\mathcal{B}(\Xi_c^0\rightarrow \Sigma^+ \pi^-)}{f_{SCS}}
        - \dfrac{\mathcal{B}(\Xi_c^0\rightarrow p \pi^-)}{f_{DCS}}$
        & $(2.00\pm 0.44)\cdot 10^{-3}$ \\[-16pt] & \\
        \hline
        \hline
    \end{tabular}
    \caption{Predictions for branching ratios or linear combinations of branching ratios obtained from next-to-leading-order $U$-spin sum rules, Eqs.~(\ref{prediction:DPV}, \ref{prediction:charged_charm}, \ref{prediction:neutral_charm-}, \ref{prediction:neutral_charm+}). All results are valid up to second order in $U$-spin breaking.}
    \label{tab:NLO_predictions-sum}
\end{table}

\paragraph{The validity of the $U$-spin expansion.}
The results presented here can be used to examine the validity of the $U$-spin expansion. As discussed in the introduction, the expansion parameter is not unambiguously defined: the numerator is fixed by the quark-mass difference, while the scale in the denominator is not known, see Eq.~\eqref{eq:epsilon_sch}. It is therefore not clear whether the expansion should be trusted.

When compared to data, the sum rules at first order show mixed behavior: some are well satisfied, while others show large deviations from the symmetry limit. By contrast, in the three examples where the second order sum rules can be tested, they are all satisfied rather well. Since the sum rules at first and second order have different forms, a direct comparison is not straightforward. Nevertheless, we believe that the current data suggests a pattern in which large deviations can appear in the leading sum rules, while the sum rules at second order remain well behaved. This observation does not prove the validity of the $U$-spin expansion, but it points in that direction. More data and further theoretical analysis are needed to test this interpretation.

\paragraph{Straightforward extensions.} We point out that it is straightforward to extend the results presented here to other phenomenologically interesting cases. For example, in systems where some decay channels are kinematically forbidden, the second-order master sum rule can be used to set bounds on the allowed rates. This is, for example, the case for $D^0 \rightarrow P^+ P^+ P^- P^-$ decays, where the $D^0 \to K^+ K^+ K^- K^-$ mode is kinematically forbidden. It is, however, not clear how robust such bounds are, as near the boundary of phase space one expects enhanced symmetry breaking and correspondingly larger theoretical corrections. Nevertheless, we believe such bounds are worth investigating.

Another possible phenomenological application of the results presented in this work involves systems that include strong decays. In cases where some initial states of the system decay weakly while others are overshadowed by strong decays, a direct experimental test of the weak modes of the latter is practically impossible. For such systems, second-order sum rules can be used to estimate the weak decay widths, providing indirect information on otherwise inaccessible processes. Additionally, the results of Section~\ref{sec:theory} apply directly to systems where all decays are strong and can therefore be used to obtain nontrivial second-order relations among strong decay rates.

\paragraph{Future work.} There are several directions we plan to pursue in follow-up work. First, the method developed here does not guarantee that all possible sum rules are obtained, nor does it provide a clear path for extending the analysis beyond second order. A detailed and systematic framework for deriving all rate sum rules for arbitrary systems, to all orders in the symmetry breaking, will be presented in a future publication~\cite{Gavrilova:inpreparation}.

Additionally, in this work we take an agnostic approach to the theoretical uncertainties. While we quote $\mathcal{O}(\varepsilon)$ and $\mathcal{O}(\varepsilon^2)$ as our parametric expectations for first- and second-order corrections, we do not attempt to quantify them. A more quantitative theoretical understanding of the size and systematics of these corrections is essential for interpreting the measurements and for assessing the potential of sum rules as probes of new physics. We plan to address this issue in future work as well.

Finally, another direction is to investigate possible dynamical explanations for the experimentally observed deviations from the sum rules. In particular, can we explain the pattern of deviations from the first order sum rules, and how is this pattern related to the corresponding deviations observed at second order?

\paragraph{Outlook.} When comparing the second-order sum rules with available data, we observe some intriguing features. The data suggest that the deviations of the second-order sum rules from the symmetry limit do not simply scale as the square of the first-order ones. In fact, the second-order sum rules appear to be satisfied more accurately than one might naively expect given the order-one deviations observed in the symmetry limit relations. It is premature to draw firm conclusions, but if these patterns are confirmed by future measurements, they may point to an underlying dynamical mechanism. We find this possibility particularly interesting, as it may help shed light on the dynamics of charm decays.

To conclude, we call for dedicated measurements of the listed charmed decay systems. We also raise the question of whether the observed features admit a dynamical interpretation and encourage further theoretical work to identify additional systems in charm physics and beyond that exhibit higher-order sum rules analogous to the ones presented in this work.

\acknowledgments

We thank Ryan Plestid for many stimulating discussions and pointing us to the literature on the Shmushkevich method. We are grateful to Livia Kong for sharing the code used to verify the sum rules. We thank Mark Wise for useful discussions.  We thank John Preskill for suggesting the phrase ``One Sum To Rule Them All,'' which inspired the title of this paper. MG thanks the CERN Theory Group for warm hospitality during a visit through the CERN Visitor Program, where part of this work was conceived and carried out. This material is based upon work supported by the U.S. Department of Energy, Office of Science, Office of High Energy Physics, under Award Number DE-SC0011632. YG is supported by the NSF grant PHY-2309456. S.S.~is supported by the STFC through an Ernest Rutherford Fellowship under reference ST/Z510233/1 and the grant ST/X003167/1.

\appendix

\section{Derivation of the Shmushkevich equation}\label{app:shmu}

In this Appendix we review the Shmushkevich method~\cite{Shmushkevich:1955,DushinShmushkevich:1956}. A systematic presentation can be found in Ref.~\cite{PinskiMacfarlaneSudarshan:1965}, where the method is discussed for isospin symmetry in strong-interaction transitions. Here we present the derivation of the method along the lines of Ref.~\cite{PinskiMacfarlaneSudarshan:1965} for systems in which the external state particles and the effective transition Hamiltonian transform as arbitrary irreducible representations of a general \(SU(2)_F\) symmetry. This extension is straightforward, and we include it for completeness.

Consider a general system of processes in which the initial state, the final state, and the effective Hamiltonian transform as arbitrary \(SU(2)_F\) irreps \(I_1,\, I_2,\,\dots,\,I_r\). As shown in Appendix~D of Ref.~\cite{Gavrilova:2022hbx}, moving irreps between the initial state, the final state, and the Hamiltonian does not affect the structure of sum rules. Therefore, without loss of generality, we can consider a system in which the transition is induced by an \(SU(2)_F\) singlet operator \(H_0^0\) between a singlet initial state and a final state carrying all nontrivial irreps,
\begin{equation}\label{app:shum:in_f_states}
    \ket{i}=\ket{0,0},\qquad\ket{f}=\ket{I_1,m_1;I_2,m_2;...;I_r,m_r}\;,
\end{equation}
where \(I_i\) are the \(SU(2)_F\) irreps describing the system and \(m_i\) their corresponding \(m\)-QNs.

Any amplitude of the system can be labeled by the set of $r$ $m$-QNs: $A(m_1,...,m_r)$, such that:
\begin{equation}\label{app:shmu:eq:m_sum}
    m_1+...+m_r=0\;.
\end{equation}
We now perform the Clebsch-Gordan decomposition of the final state $\ket{f}$, in Eq. (\ref{app:shum:in_f_states}), which is given by the following tensor product
\begin{equation}\label{app:shum:eq:f_product}
    \ket{f}= \ket{I_1,m_1}\otimes\cdots\otimes\ket{I_r,m_r}\;.
\end{equation}
To decompose \(\ket{f}\) into irreducible \(SU(2)_F\) representations, we choose a tensor product ordering in which the final state irreps are combined sequentially,
\begin{equation}
    (((I_1\otimes I_2)\otimes I_3)\otimes \cdots)\otimes I_r\,,
\end{equation}
and we introduce the intermediate quantum numbers \(J_1,J_2,\ldots,J_{r-1}\), which, by definition, can have the following values:
\begin{align}
    J_1\in\{|I_1-I_2|,...,I_1+I_2\},\\
    J_2\in\{|J_1-I_3|,...,J_1+I_3\},\nonumber\\
    \cdots \nonumber\\
    J_{r-1}\in\{|J_{r-2}-I_{r}|,...,J_{r-2}+I_r\}\nonumber\;.
\end{align}
Also, we introduce the following notation for the Clebsch-Gordan coefficients 
\begin{equation}
    C^{J M}_{\substack{I_1 m_1\\ I_2 m_2}} \equiv \bra{J,M}\ket{I_1,m_1;I_2,m_2}\;.
\end{equation}
Hence, the product state in Eq.~\eqref{app:shum:eq:f_product} can be expanded by the Clebsch-Gordan decomposition as
\begin{equation}
    \ket{f}=\sum_{J_1,...,J_{r-1}} C^{J_1 M_1}_{\substack{I_1 m_1\\ I_2 m_2}}\times C^{J_2 M_2}_{\substack{J_1 M_1\\ I_3 m_3}}\times...\times C^{J_{r-1} M_{r-1}}_{\substack{J_{r-2} M_{r-2}\\ I_r m_r}} \;\ket{J_1,...,J_{r-1},M_{r-1}}\;,
\end{equation}
where $M_j=\sum_{k=1}^{j+1}m_k$, and where $M_{r-1}=m_1+...+m_r=0$ by Eq.~(\ref{app:shmu:eq:m_sum}). As a consequence, we find the following decomposition for the amplitudes of the processes in the symmetry limit

\begin{align}\label{App:eq:Wigner}
\hat{A}(m_1,...,m_r)=\bra{i}H_0^0\ket{f}=\bra{0}H_0^0 \ket{I_1,m_1;I_2,m_2;...;I_r,m_r} \\
= \sum_{J_1,\ldots,J_{r-1}} C^{J_1 M_1}_{\substack{I_1 m_1\\ I_2 m_2}}\times C^{J_2 M_2}_{\substack{J_1 M_1\\ I_3 m_3}}\times...\times C^{J_{r-1} M_{r-1}}_{\substack{J_{r-2} M_{r-2}\\ I_r m_r}}\;X(J_1,...,J_{r-1})\,\nonumber ,
\end{align}
where the matrix elements $X$ only depend on the values of the quantum numbers $J_1,...J_{r-1}$, and are called reduced matrix elements. By  conservation of the $SU(2)_F$ angular momentum (the Wigner-Eckart theorem), we also have $J_{r-1}=0$.

We define the quantity:
\begin{equation}\label{App:eq:sigma}
    T_i(m_i)=\sum_{\{m_j\}_{j\neq i}}|\hat{A}(m_1,...,m_r)|^2\;.
\end{equation}
We also consider the orthogonality relations of the Clebsch-Gordan coefficients, Ref.~\cite{PinskiMacfarlaneSudarshan:1965},
\begin{align}
    \sum_{m_1,m_2}C^{J M}_{\substack{I_1 m_1\\ I_2 m_2}} C^{J' M'}_{\substack{I_1 m_1\\ I_2 m_2}} &= \delta_{J J'}\delta_{M M'}\;,\\
    \sum_{m_1,M}C^{J M}_{\substack{I_1 m_1\\ I_2 m_2}} C^{J M}_{\substack{I_1 m_1\\ I'_2 m'_2}} &= \frac{2J+1}{2I_2+1}\delta_{I_2 I_2'}\delta_{m_2 m_2'}\;.
\end{align}
If we plug Eq.~\eqref{App:eq:Wigner} into Eq.~\eqref{App:eq:sigma}, by summing over all the $m-$QNs, except the fixed one $m_i$, the orthogonality relations will make the $m_i$ dependence disappear, and we find:
\begin{equation}
    T_i(m_i) = \sum_{J_1,...,J_{r-1}} \frac{2J_{r-1}+1}{2I_i+1}|X(J_1,...,J_{r-1})|^2
\end{equation}
which tells us that $T_i(m_i)$ does not depend on $m_i$. As a consequence,
\begin{equation}\label{App:eq:Master}
    T_i(-I_i)=T_i(-I_i+1)=...=T_i(+I_i)\;.
\end{equation}
Given the definition of \(T_i(m_i)\) in Eq.~(\ref{App:eq:sigma}), Eq.~(\ref{App:eq:Master}) gives a set of sum rules among amplitudes squared. These relations are derived in the symmetry limit and are independent of the underlying dynamics: they follow solely from \(SU(2)_F\) symmetry.

In conclusion, the generic observables \(\hat{\sigma}(m_1,\ldots,m_r)\) introduced in Section~\ref{sec:theory} are given by squared amplitudes convoluted with appropriate kinematic functions, \(\hat{\sigma}\propto |\hat{A}|^2\). In the symmetry limit, all particles within a given \(SU(2)_F\) irrep \(I_i\) are mass-degenerate; therefore, all channels in the system share the same kinematic factors. Consequently, Eq.~(\ref{App:eq:Master}) can be rewritten in terms of the inclusive observables \(\hat{\sigma}_i(m_i)\) defined in Eq.~(\ref{eq:sigma_i_def}), yielding
\begin{equation}\label{app:Master}
    \hat{\sigma}_i(-I_i)=\hat{\sigma}_i(-I_i+1)=\cdots=\hat{\sigma}_i(+I_i)\;.
\end{equation}
Note that we implicitly assume here that all irreps in the system are distinguishable. The case with identical irreps is discussed in Section~\ref{sec:identical}.

\section{$D_q^-\!\to P^+P^-P^-$: an explicit example with identical multiplets}
\label{app:identical_DPPP}

In this Appendix we illustrate the treatment of systems with identical multiplets discussed in Section~\ref{sec:identical} using, as an example, the system of three-body decays
$D_q^-\!\to P^+P^-P^-$ that we also consider in Section~\ref{sec:DtoPPP}. The $U$-spin doublets $P^\pm$ are defined in Eq.~\eqref{eq:Ppm_def}, and $D_q^-$ is defined in Eq.~\eqref{eq:D_ds_def}, the full list of decays in this system is given in Table~\ref{tab:DPPP}. This system contains two identical $U$-spin doublets, $P^-$, in the final state. Here we show explicitly for this system that the momentum-labeled (differential) Shmushkevich sum rule, Eq.~\eqref{eq:shmu_dif}, yields a relation of the same structure among integrated rates, Eq.~\eqref{eq:shmu_phys}, without additional combinatorial factors.

\paragraph{Sum rules between amplitudes squared.} 

As discussed in Section~\ref{sec:identical} and shown in Section~\ref{app:shmu}, the Shmushkevich sum rules, Eq.~\eqref{eq:shmu_master}, are formally derived for systems in which all irreps are distinguishable, and they take the form of relations among squared CKM-free amplitudes $\hat{A}(i\to f)$, see Eq.~\eqref{App:eq:Wigner} and the derivation that follows. We define 
\begin{equation}\label{eq:A2_CKM_free}
    |\hat{A}(i\to f)|^2 \equiv \frac{|A(i\to f)|^2}{f_\text{CKM}}\,,
\end{equation}
where $A(i\to f)$ is the physical decay amplitude and, for weak charm decays, $f_\text{CKM}$ is given in Eq.~\eqref{eq:f_CKM}. For the rest of this Appendix we refer to the CKM-free amplitudes $\hat{A}(i\to f)$ simply as amplitudes.

Since the amplitudes are functions of the final-state momenta, and in the symmetry limit all channels of a $U$-spin system share identical kinematics, these sum rules must be interpreted as relations among amplitudes squared, where all amplitudes are evaluated at the same kinematic point.
Except on a set of measure zero where the two $P^-$ momenta coincide exactly (which does not affect phase space integrals), at a fixed point in phase space the irreps are effectively distinguishable, because one can assign a momentum label to each irrep. In the case of $D_q^-\to P^+ P^- P^-$, we therefore write the irreps in the final state as
\begin{equation}
P^+_{q}\,,\qquad P^-_{p_1}\,,\qquad P^-_{p_2}\,,
\end{equation}
so that the two copies of $P^-$ are distinguished by their momenta, $p_1\neq p_2$.

In this description, the Shmushkevich equation, Eq.~\eqref{eq:shmu_master}, applies directly and yields a relation among momentum-labeled squared amplitudes, Eq.~\eqref{eq:shmu_dif}. Recall that the Shmushkevich equations in Sections~\ref{sec:shmu}--\ref{sec:sym+shmu=love} apply to any $\hat{\sigma}\propto|\hat{A}|^2$, with squared amplitudes themselves providing one example. Since the group-theoretical structure of $D_q^-\!\to P^+P^-P^-$ contains only one higher irrep---the triplet in the Hamiltonian, $I_H=1$ (as in all weak charm-decay systems, see Section~\ref{sec:master})---there is a single nontrivial Shmushkevich equation of the form~\eqref{eq:shmu_dif}. It can be written as an equality between inclusive sums, Eq.~\eqref{eq:sigma_i_def}, of squared amplitudes for different values of the $m$-QN $m_H$ of the Hamiltonian. As shown in Section~\ref{sec:charm}, $m_H=+1,-1,0$ correspond to DCS, CF, and SCS decays, respectively. We thus write the inclusive sums entering Eq.~\eqref{eq:shmu_dif}, for the triplet $I_H=1$, as 
\begin{align}
\hat{\sigma}^{\text{diff}}_H(+1)=&
\big|\hat{A}(D^- \to \pi^+_{q}\,K^-_{p_1}\,\pi^-_{p_2})\big|^2
+\big|\hat{A}(D^- \to \pi^+_{q}\,\pi^-_{p_1}\,K^-_{p_2})\big|^2 \nonumber\\
+&\big|\hat{A}(D^- \to K^+_{q}\,K^-_{p_1}\,K^-_{p_2})\big|^2
+\big|\hat{A}(D_s^- \to \pi^+_{q}\,K^-_{p_1}\,K^-_{p_2})\big|^2\,,
\qquad &\text{(DCS)} \label{eq:sigmaH_dif_p1}
\\[2pt]
\hat{\sigma}^{\text{diff}}_H(-1)= &
\big|\hat{A}(D_s^- \to K^+_{q}\,\pi^-_{p_1}\,K^-_{p_2})\big|^2
+\big|\hat{A}(D_s^- \to K^+_{q}\,K^-_{p_1}\,\pi^-_{p_2})\big|^2 \nonumber\\
+&\big|\hat{A}(D_s^- \to \pi^+_{q}\,\pi^-_{p_1}\,\pi^-_{p_2})\big|^2
+\big|\hat{A}(D^- \to K^+_{q}\,\pi^-_{p_1}\,\pi^-_{p_2})\big|^2\,,
\qquad &\text{(CF)} \label{eq:sigmaH_dif_m1}
\\[2pt]
\hat{\sigma}^{\text{diff}}_H(0)= {}&
\big|\hat{A}(D^- \to K^+_{q}\,K^-_{p_1}\,\pi^-_{p_2})\big|^2 + \big|\hat{A}(D_s^- \to \pi^+_{q}\,\pi^-_{p_1}\,K^-_{p_2})\big|^2 \nonumber\\
+&\big|\hat{A}(D^- \to K^+_{q}\,\pi^-_{p_1}\,K^-_{p_2})\big|^2 + \big|\hat{A}(D_s^- \to \pi^+_{q}\,K^-_{p_1}\,\pi^-_{p_2})\big|^2 \nonumber\\
+&\big|\hat{A}(D^- \to \pi^+_{q}\,\pi^-_{p_1}\,\pi^-_{p_2})\big|^2 + \big|\hat{A}(D_s^- \to K^+_{q}\,K^-_{p_1}\,K^-_{p_2})\big|^2\,,
\qquad &\text{(SCS)} \label{eq:sigmaH_dif_0}
\end{align}
where we keep the momentum dependence of the amplitudes explicit. The symmetry limit Shmushkevich sum rules, Eq.~\eqref{eq:shmu_dif}, are then written as
\begin{equation}\label{app:eq:shmu_H}
    \hat{\sigma}^\text{diff}_H(+1) = \hat{\sigma}^\text{diff}_H(0) =\hat{\sigma}^\text{diff}_H(-1)\,,
\end{equation}
and they hold point-by-point in phase space.

\paragraph{Sum Rules between integrated decay rates.}
Next, we integrate the sum rule in Eq.~\eqref{app:eq:shmu_H} over the three-body phase space. In the symmetry limit the phase space measure is the same for all channels in the system, so we write
\begin{equation}\label{app:eq:shmu_int}
    \int d\Pi_3\,\hat{\sigma}^{\text{diff}}_H(+1)
    = \int d\Pi_3\,\hat{\sigma}^{\text{diff}}_H(0)
    = \int d\Pi_3\,\hat{\sigma}^{\text{diff}}_H(-1)\,.
\end{equation}
As discussed in Section~\ref{sec:identical}, upon integration there are two sources of combinatorial factors.

\emph{(i) Two identical particles in the final state.}
If the final state contains two identical mesons $\phi\phi$ ($\pi^-\pi^-$ or $K^-K^-$), then the physical (CKM-free) decay rate includes a factor $1/2!$,
\begin{equation}
\hat{\Gamma}(i\to \cdots\phi\phi)=\frac{1}{2}\int d\Pi_3\,
|\hat{A}(i\to \cdots \phi_{p_1}\,\phi_{p_2})|^2\,,
\end{equation}
and thus, the momentum-labeled integrals of squared amplitudes for processes with identical particles in the final state appearing in Eqs.~\eqref{eq:sigmaH_dif_p1}-\eqref{eq:sigmaH_dif_0} map upon integration to
\begin{equation}
\int d\Pi_3\,|\hat{A}(i\to\cdots \phi_{p_1}\,\phi_{p_2})|^2
=2\,\hat{\Gamma}(i\to\cdots\phi\phi)\,.
\label{eq:DPPP_map_identical}
\end{equation}

\emph{(ii) Two different particles with exchanged momentum labels.}
If the final state contains two different mesons $\phi\chi$ ($K^-\pi^-$), then both momentum orderings for these mesons ($K^-_{p_1}\pi^-_{p_2}$ and $\pi^-_{p_1}K^-_{p_2}$) integrate to the same physical (CKM-free) decay rate,
\begin{equation}
\int d\Pi_3\,\abs{A(i\to\cdots \phi_{p_1}\,\chi_{p_2})}^2
=\int d\Pi_3\,\abs{A(i\to\cdots \chi_{p_1}\,\phi_{p_2})}^2
=\Gamma(i\to\cdots\phi\chi)\,.
\end{equation}
Consequently, whenever the sums in Eqs.~\eqref{eq:sigmaH_dif_p1}-\eqref{eq:sigmaH_dif_0} contain both orderings, their combined contribution upon integration is
\begin{equation}
\int d\Pi_3\Big(\abs{A(i\to\cdots \phi_{p_1}\,\chi_{p_2})}^2
+\abs{A(i\to\cdots \chi_{p_1}\,\phi_{p_2})}^2\Big)
=2\,\Gamma(i\to\cdots\phi\chi)\,.
\label{eq:DPPP_map_distinguishable}
\end{equation}

Applying Eqs.~\eqref{eq:DPPP_map_identical} and~\eqref{eq:DPPP_map_distinguishable} to
Eqs.~\eqref{eq:sigmaH_dif_p1}--\eqref{eq:sigmaH_dif_0} and integrating over $d\Pi_3$, we obtain
\begin{align}
\int d\Pi_3\,\hat{\sigma}^{\text{diff}}_H(+1) \;=\;&
2\Big[
\hat{\Gamma}(D^- \to \pi^+K^-\pi^-)
+\hat{\Gamma}(D^- \to K^+K^-K^-)
\nonumber\\[-2pt]
+&\hat{\Gamma}(D_s^- \to \pi^+K^-K^-)
\Big]\,, \label{eq:DPPP_sigmaH_map_p1}
\\[4pt]
\int d\Pi_3\,\hat{\sigma}^{\text{diff}}_H(-1) \;=\;&
2\Big[
\hat{\Gamma}(D_s^- \to K^+K^-\pi^-)
+\hat{\Gamma}(D_s^- \to \pi^+\pi^-\pi^-)
\nonumber\\[-2pt]
+&\hat{\Gamma}(D^- \to K^+\pi^-\pi^-)
\Big]\,, \label{eq:DPPP_sigmaH_map_m1}
\\[4pt]
\int d\Pi_3\,\hat{\sigma}^{\text{diff}}_H(0) \;=\;&
2\Big[
\hat{\Gamma}(D_s^- \to \pi^+K^-\pi^-)
+\hat{\Gamma}(D^- \to K^+K^-\pi^-)
\nonumber\\[-2pt]
+&\hat{\Gamma}(D_s^- \to K^+K^-K^-)
+\hat{\Gamma}(D^- \to \pi^+\pi^-\pi^-)
\Big]\,. \label{eq:DPPP_sigmaH_map_0}
\end{align}
Thus, when we express the sum rule in Eq.~\eqref{app:eq:shmu_int} in terms of the physical CKM-free rates, the factors of $2$ cancel, and we obtain an equation of the form in Eq.~\eqref{eq:shmu_phys}\,.

\section{Including phase space effects in two-body decays}\label{app:two-body}

In Section~\ref{sec:examples}, we tested the second-order charm master sum rule, Eq.~\eqref{eq:RH}, in the form of relations among CKM-free rates $\hat{\Gamma}$. We note that, formally, the same relations also hold to second order when written in terms of CKM-free squared amplitudes, Eq.~(\ref{eq:A2_CKM_free}). That is, including or omitting phase-space effects amounts to corrections to the master sum rule that are theoretically second order in the breaking. This can be seen by noting that the general argument for the validity of the sum rules to second order relies on the fact that the sum rules are invariant under exchange of $U$-spin conjugated particles, a property that holds for both the rates and the squared amplitudes. Thus, the second-order charm master sum rule can also be written in terms of CKM-free amplitudes squared:
\begin{equation}\label{eq:A2_master}
    R_{A}(\text{$U$-spin system})=\frac{\sum_{\text{CF,\,DCS}}|\hat{A}|^2}
         {\sum_{\text{SCS}}|\hat{A}|^2}
    = 1 + \mathcal{O}\left(\varepsilon^2\right) \, .
\end{equation}

For two-body decays, the phase-space effects are reduced to a simple multiplicative factor, as
the final-state kinematics is fully determined. Then, CKM-free amplitudes squared can be extracted from measured rates by factoring out the two-body phase space:
\begin{equation}
    |\hat{A}(i\rightarrow f)|^2= \hat{\Gamma}(i\rightarrow f)\frac{8\pi M_i^2}{p_f}
    =\frac{\mathcal{B}(i\to f)}{\tau_i\, f_\text{CKM}}\frac{8\pi M_i^2}{p_f}\,,
\end{equation}
where $M_i$ is the mass of the decaying particle and $p_f$ is the center-of-mass momentum of the final state. Thus, for two-body decays, Eq.~\eqref{eq:A2_master} can also be tested against experimental data.

\begin{table}[t]
    \centering
    \begin{tabular}{|l|c|}
        \hline
        \hline
        Particle & Mass [MeV] \\
        \hline
        $K^{+}$       & $493.677\pm0.015$ \\
        $\pi^{+}$     & $139.57039\pm0.00018$ \\
        $\rho^{+}$    & $775.11\pm0.34$ \\
        $K^{*+}$      & $891.67\pm0.26$ \\
        \hline
        \hline
    \end{tabular}
    \caption{Final-state particle masses \cite{PDG:2024cfk}. For the $\rho^+$ we use the mass determined in $e^+e^-\rightarrow \pi^+\pi^-\pi^0$ and $\tau$ decays. For the $K^{*+}$ resonance we use the mass measured in hadron production.}
    \label{tab:particle_masses}
\end{table}

We test Eq.~\eqref{eq:A2_master} for two fully measured systems of two-body decays, $D^0\to P^+P^-$ and $D^0\to P^-V^+$, discussed in Sections~\ref{sec:DtoPP} and~\ref{sec:DtoPV}, respectively. Using the inputs in Tables~\ref{tab:DPP},~\ref{tab:DPV}, and~\ref{tab:particle_masses}, we find 
\begin{equation}
    R_A(D^0\rightarrow P^+P^-)=
    \frac{|\hat{A}(D^0 \to K^+ \pi^-)|^2 + |\hat{A}(D^0 \to \pi^+ K^-)|^2}
         {|\hat{A}(D^0 \to K^+ K^-)|^2 + |\hat{A}(D^0 \to \pi^+ \pi^-)|^2}
    = 0.80 \pm 0.01\,, 
\end{equation}
\begin{equation}
    R_A(D^0\rightarrow P^-V^+)=
    \frac{|\hat{A}(D^0\rightarrow K^-\rho^+)|^2+|\hat{A}(D^0\rightarrow \pi^-K^{*+})|^2}
         {|\hat{A}(D^0\rightarrow K^-K^{*+})|^2 + |\hat{A}(D^0\rightarrow \pi^-\rho^+)|^2}
    = 0.91\pm0.18\;. 
\end{equation}

Comparing $R_A(D^0 \to P^+P^-)$ above with $R_H(D^0 \to P^+P^-)$ in Eq.~\eqref{eq:master_DPP}, we see that
$R_A(D^0 \to P^+P^-)$ lies further from the symmetry limit than
$R_H(D^0 \to P^+P^-)$. This is somewhat surprising, as the phase-space
factor introduces an additional source of symmetry breaking, and thus,
upon removing it, one might naively expect the result to move closer to
the symmetry limit. However, this need not be the case, since different
sources of symmetry breaking can shift the result in different
directions.

Comparing $R_A(D^0 \to P^-V^+)$ above with $R_H(D^0 \to P^-V^+)$ in Eq.~\eqref{eq:RH-DPV}, we see that they
are compatible with each other, albeit with large uncertainties. As more data becomes available, it will be interesting to see how these ratios evolve and how they compare for other two-body systems.

\bibliographystyle{JHEP}
\bibliography{biblio.bib}

\end{document}